\documentclass[pdflatex,sn-mathphys]{sn-jnl}

\usepackage{xr}
\usepackage{amsmath}
\usepackage{lineno}

\makeatletter
\newcommand*{\addFileDependency}[1]{
  \typeout{(#1)}
  \@addtofilelist{#1}
  \IfFileExists{#1}{}{\typeout{No file #1.}}
}
\makeatother

\newcommand*{\myexternaldocument}[1]{%
    \externaldocument{#1}%
    \addFileDependency{#1.tex}%
    \addFileDependency{#1.aux}%
}

\myexternaldocument{sn-si}

\renewcommand{\eqref}[1]{Eq. \textup{(\ref{#1})}}

\jyear{2022}%

\theoremstyle{thmstyleone}%
\theoremstyle{thmstyletwo}%

\theoremstyle{thmstylethree}%

\let\oldequation\equation
\let\oldendequation\endequation

\renewenvironment{equation}
  {\linenomathNonumbers\oldequation}
  {\oldendequation\endlinenomath}

\newcommand*{\defeq}{\stackrel{\text{def}}{=}}

\begin{document}

\title[How networks shape diversity for better or worse]{How networks shape diversity for better or worse}


\author[1,2]{\fnm{Andrea} \sur{Musso}}\email{andrea.musso@gess.ethz.ch}
\author[1,2]{\fnm{Dirk} \sur{Helbing}}\email{dirk.helbing@gess.ethz.ch}
\affil[1]{\orgdiv{Computational Social Science}, \orgname{ETH Z{\"u}rich}, \orgaddress{\street{Ramistrasse 81}, \city{Z{\"u}rich}, \postcode{8092}, \country{Switzerland}}}

\affil[2]{\orgdiv{Complexity Science Hub}, \orgaddress{\street{Josefst{\"a}dter Strasse 39}, \city{Vienna}, \postcode{1080}, \country{Austria}}}


\abstract{Socio-diversity, the variety of human opinions, ideas, behaviors and styles, has profound implications for social systems. While it fuels innovation, productivity, and collective intelligence, it can also complicate communication and erode trust. So what mechanisms can influence it? This paper studies how fundamental characteristics of social networks can support or hinder socio-diversity. It employs models of cultural evolution, mathematical analysis, and numerical simulations. We find that pronounced inequalities in the distribution of connections obstruct socio-diversity. In contrast, the prevalence of close-knit communities, a scarcity of long-range connections, and a significant tie density tend to promote it. These results open new perspectives for understanding how to change social networks to sustain more socio-diversity and, thereby, societal innovation, collective intelligence, and productivity.}

\keywords{networks, diversity, cultural evolution, opinion dynamics, graph theory}



\maketitle
\section*{Introduction}
In his seminal work, \emph{The Selfish Gene} \cite{dawkins1990selfish}, Richard Dawkins proposes a compelling perspective on culture and its evolution. He argues that, by viewing \emph{memes}\footnote{The memes considered in this paper should not be confused with the homonymous ``memes'' circulating on social media. Social media``memes'' are funny internet content, while this paper's ``memes'' are the building blocks of culture.}---a collective term encompassing human ideas, opinions, behaviors, and styles---as cultural counterparts to genes, one can use the foundational principles of biological evolution to explain the origins and development of cultural traits. 

Biological evolution is driven by mutation and selection, acting on biology's fundamental units: genes \cite{nowak2006evolutionary}. Mutation introduces new genetic variants into a population, which undergo selection through individual interactions. The nature of these interactions varies: some individuals engage with each other frequently, while others are predominantly isolated. Such patterns of selective interaction, commonly known as the interaction network structure, critically determine survival chances, thereby influencing which genetic variants are perpetuated through reproduction and which become extinct. In essence, the structure of the interaction network is instrumental in shaping the variety of genes present in the population, or, in other words, its bio-diversity. 

Research suggests a complex relationship between the structure of interaction networks and bio-diversity. For example, certain inter-individual interaction networks can increase the advantage of fitter individuals, potentially reducing bio-diversity  \cite{nowak2006evolutionary, lieberman2005evolutionary}. In contrast, the pronounced nestedness found in inter-species interaction networks seems to boost bio-diversity by reducing direct competition among species \cite{bastolla2009architecture}.

Socio-diversity, defined as the variety of memes (i.e., ideas, opinions, behaviors, and styles) present in a society, is a social parallel to bio-diversity. 
Dawkins' perspective on cultural evolution suggests that socio-diversity emerges from the interplay of imitation and innovation, acting upon culture’s basic units: memes \cite{boyd1988culture, simmel1957fashion}. 
Innovation, akin to biological mutation, creates new cultural variants (new memes). Imitation, as counterpart of biological selection, determines which variants diffuse across society. 

Just as the structure of species interaction networks influences bio-diversity, the structure of social interaction networks affects socio-diversity. There is substantial evidence that the network structure in which individuals are embedded \cite{granovetter1985economic} significantly influences the diffusion of various memes, such as obesity \cite{christakis2007spread}, smoking \cite{christakis2008collective}, cooperation \cite{hauert2004spatial, nowak2006evolutionary, nowak2006five}, and product adoption \cite{granovetter1978threshold}. Generally, clustered networks, characterized by numerous closed triangles (i.e., your friends are also friends), excel at spreading memes, such as complex behavior, requiring repeated endorsement \cite{centola2010spreada, ugander2012structurala, granovetter1978threshold}. The dense local connections in these networks provide the repeated exposure necessary for such memes to take hold. Conversely, networks characterized by an abundance of long-range connections are more effective at disseminating simpler memes that need minimal reinforcement \cite{onnela2007structure, granovetter1973strength}. These far-reaching connections enable quick meme transmission across diverse network areas, facilitating rapid spread. 

Network structure also significantly shapes meme creation. Efficient networks, characterized by short distances between nodes, appear to hinder the creation of radically novel memes by facilitating blind imitation \cite{lazer2007network, mason2012collaborative}. In contrast, networks with larger average distances between nodes, seem to foster the generation of novelty by providing fewer imitation opportunities \cite{lazer2007network, mason2012collaborative}. 

In summary, while there is substantial evidence on the influence of network structure on patterns of meme diffusion and creation, and these patterns clearly affect socio-diversity, research directly examining the relationship between network structure and socio-diversity is scarce. As a result, a systematic understanding of this relationship is still lacking, especially when compared to the comprehension in ecology and conservation biology of the similar relationship between network structure and bio-diversity\footnote{This is not surprising: "Understanding the factors determining biodiversity is arguably the most fundamental problem in ecology and conservation biology" \cite{bastolla2009architecture}, in contrast to the less central role of socio-diversity studies in many social sciences}. Crucial questions remain open:  How can we measure a social network's potential to foster socio-diversity? Which social networks enhance socio-diversity, and which ones diminish it?

This paper addresses these important questions by introducing a novel index linking network structure and socio-diversity: the \emph{structural diversity index}. 
This index, derived from the theory of random walks on networks \cite{rayleigh1905problem, levin2017markov}, quantifies the propensity of a network to support socio-diversity; its ability to protect unpopular memes from being crushed by more popular ones. 
With our novel index, understanding if a social network enhances or diminishes socio-diversity becomes straightforward: it suffices to compare the network's index value against a benchmark (such as the complete network). 
If a network’s index value is higher than the benchmark, then the network promotes socio-diversity; otherwise it hinders it. 
The index is designed for scalability, can handle large networks and is easily accessible through the Python package accompanying this paper.

Employing our novel index, we conducted an extensive exploration of the relationship between network structure and socio-diversity across a broad range of real-world and synthetic networks. 
We focused on some key network characteristics, such as the shape of the degree distribution, the edge density and the prevalence of long-range connections. Our selection of these characteristics is underpinned by four guiding hypotheses. 
First, networks dominated by highly connected individuals may see diminished socio-diversity as a result of the disproportionate influence on cultural spreading exerted by such individuals. 
Second, networks with prevalent long-range connections might display lower socio-diversity due to the homogenizing effects of such ties. 
The pervasive spread of global pop culture and its consequential effect on local cultures exemplifies the possible homogenizing impact of these long-range connections \cite{appadurai1996modernity}. 
Third, networks with a high density of connections may bolster socio-diversity. 
Specifically, increasing the number of connections diminishes the influence of each individual connection and, thereby, lowers the chances of viral meme cascades \cite{granovetter1978threshold}. 
Fourth, networks characterized by numerous close-knit communities may enhance socio-diversity. 
In fact, when these communities share homogeneous memes, they limit exposure to new memes and foster the persistence of those already adopted.

We initiated our investigation by examining two prominent synthetic network models: scale-free \cite{barabasi1999emergence} and Watts-Strogatz \cite{watts1998collective} networks. 
Although these models might seem simplistic, they serve as useful tools to isolate the effect of different network characteristics. 
Specifically, scale-free networks offer a lens to study degree-heterogeneity, or in simpler terms, the unequal distribution of connections among nodes. 
Our analysis of these networks' structural diversity index suggest that such a high disparity in connections can reduce socio-diversity. 
This aligns with the hypothesis that individuals with numerous connections may inadvertently suppress socio-diversity due to their pronounced influence on cultural transmission. 
Conversely, Watts-Strogatz networks provide a framework to understand the ramifications of long-range connections and close-knit communities. 
Our findings indicate that less long-range connections and more close-knit communities are positive for socio-diversity. 
This is compatible with the idea that cultural convergence toward a ``global village'' is expected to erode socio-diversity.

Moving beyond these simplistic models, we broadened our investigation to include hundreds of real-world networks. 
Using the comprehensive real-world network database provided by graph-tool \cite{peixoto2014graphtool}, we computed the structural diversity index of networks originating from various social contexts. 
Our findings reinforce that high inequality in the distribution of connections suppresses socio-diversity, while scarcity of long-range connections and a prevalence of close-knit communities amplify it. 
Furthermore, we observed that a high density of connections also tends to improve socio-diversity.  

Although the factors shaping socio-diversity have drawn interest from social scientists \cite{axelrod1997disseminationa, weitzman1992diversity, huckfeldt2004political, stirling2007general, klemm2003global}, its role in social science research has not reached the prominence of bio-diversity in ecology and conservation biology. 
Yet, socio-diversity has profound real-world implications \cite{page2008difference}. 
On the positive side, socio-diversity is a catalyst for innovation \cite{feldman1999innovation}, promotes cooperation \cite{santos2008social, vasconcelos2021segregationa}, and can increase productivity \cite{bettencourt2014professional, gomez-lievano2016explaininga}. 
Research shows that protecting novel and rare ideas from premature dismissal \cite{lazer2007network, centola2022network} and encouraging independent thought \cite{hong2004groups,lorenz2011howa, bernstein2018how, woolley2010evidence} can enhance a group’s ability to solve all kinds of problems, making it more “intelligent”, more productive and more innovative. Conversely, socio-diversity can pose challenges to group cohesion \cite{guzzo1996teams, webber2001impact}, impede effective communication \cite{zenger1989organizational}, and erode trust \cite{glaeser2000measuring, putnam2007pluribus}. 
Our research seeks to reduce the relative disparity in attention between bio- and socio-diversity by shedding light on the interplay between network structure and socio-diversity.

\section*{Results}
\subsection*{Structural diversity index}
Consider a social network represented abstractly by a connected undirected graph $G$. 
In this representation, each vertex stands for an individual, and edges symbolize undirected and mutual relationships, such as friendships, acquaintances, or interactions.

For illustration, imagine that two individuals, Alice and Bob, decide to play the ``random social exploration game'', a variation of Milgram's celebrated small-world experiment \cite{milgram1967small}. 
In this game, Alice and Bob each randomly select a friend from their network and send them a letter. This letter carries a simple instruction: ``Please choose a friend at random and forward this letter to them." 
Every recipient follows this directive, passing the letter onward within their network. 
As the letters get forwarded again and again, they randomly explore the social circles of both Alice and Bob. 
The game ends when the two letters \emph{meet}, i.e., when they simultaneously end up in the mailbox of the same individual. 

\begin{figure}[htp]
    \centering
    \includegraphics[width=0.5\linewidth]{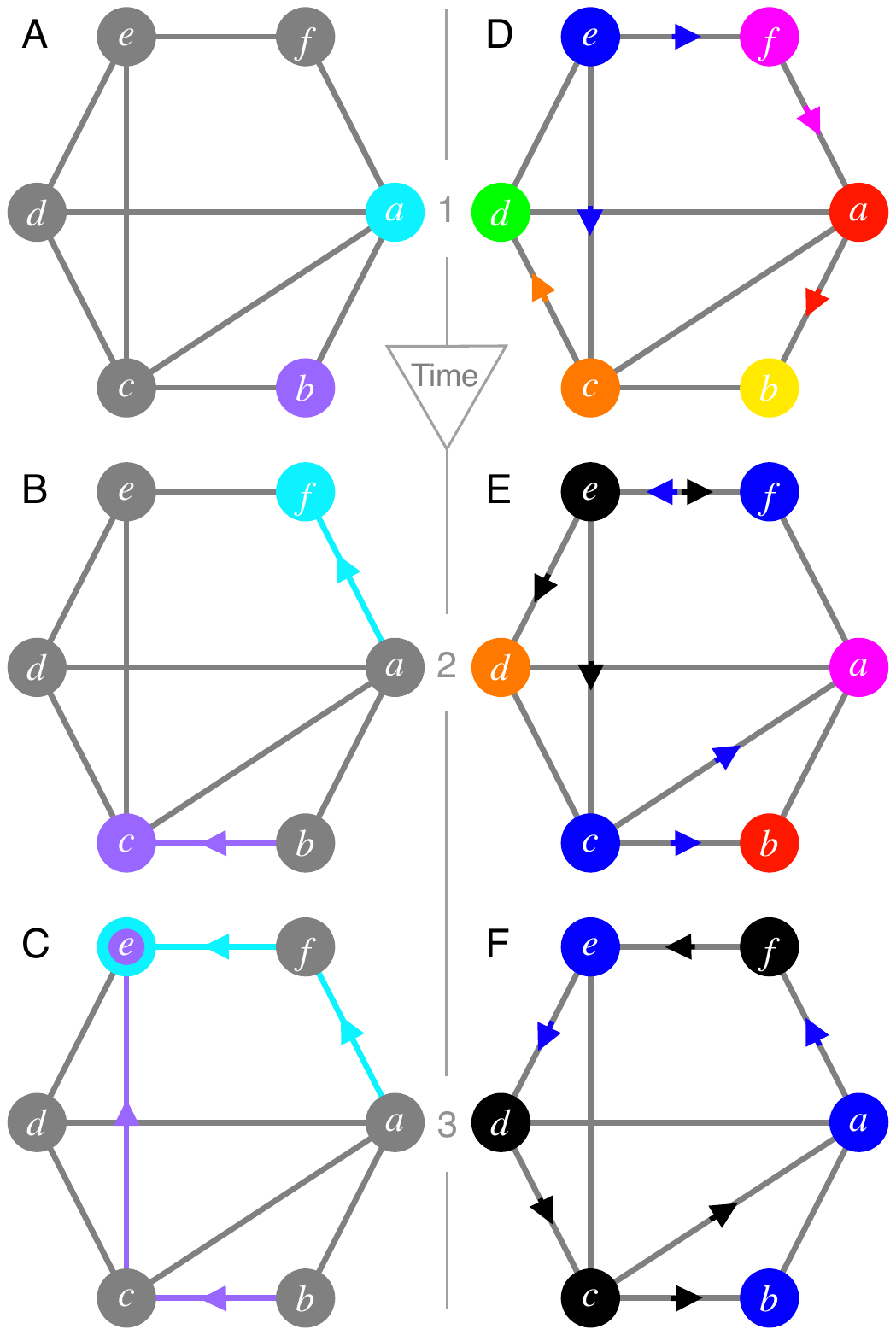}
    \caption{Illustration of the random social exploration game (A-C) and the progression of cultural evolution in a small social network (D-F). The network features six individuals: Alice ($a$), Bob ($b$), Carla ($c$), Darcy ($d$), Elon ($e$), and Frank ($f$). \textbf{(A-C)} In the random social exploration game, letters sent by Alice and Bob are randomly forwarded through the network until they meet in a mailbox (see main text for details). In this illustration, the letters' journeys are marked by highlighted edges, converging at Elon's mailbox. \textbf{(D-F)} These panels illustrate cultural evolution within our social network in three time steps. Each panel depicts both the present meme distribution and its imminent evolution by colors. An individual's current meme is reflected by the color of their vertex, whereas upcoming changes are represented by the color and direction of arrows pointing toward their vertex. For instance, Alice's red meme in Panel D, transforms into Frank's magenta meme (see Panel E), as indicated by the magenta arrow pointing from Frank to Alice. It is noteworthy that in Panel D, no arrows point towards Elon, suggesting he does not mimic others at $t=1$, but instead introduces a new meme (the black meme).}
    \label{figure:illustration_vm}
\end{figure}

As an example, consider the hexagonal-shaped social network depicted in Figure \ref{figure:illustration_vm}A. 
This network comprises six individuals: Alice ($a$), Bob ($b$), Carla ($c$), Darcy ($d$), Elon ($e$) and Frank ($f$). 
In the initial phase of the random social exploration game (shown in Panel A), Alice selects a friend at random to send her letter. 
Thus, her letter stands an equal chance of landing with Bob, Carla, Darcy, or Frank. 
In this particular illustration, fate dictates the letter to be sent to Frank, as shown in Panel B. 
Upon receiving the letter, Frank, too, makes a random choice, deciding to forward the letter to Elon. Simultaneously, Bob's letter also finds its way to Elon, having first been relayed through Carla. 
It is at this point, in Elon's mailbox, that the letters from Alice and Bob \emph{meet}, marking the game's end.

The \emph{expected meeting time} of the network $G$, denoted $\langle M_G \rangle$, is defined as the average number of forwards required for the letters to meet. 
Specifically, it is computed by conducting many repetitions of the random social exploration game, with letters starting out from different individuals and then averaging the number of forwards necessary for the letter to meet in each game iteration. 
Formally, the expected meeting time $\langle M_G \rangle$ is defined as the average number of steps before two uniformly started random walks on $G$ visit the same vertex simultaneously. 
It is a well-studied network statistic \cite{aldous1991meeting, kanade2023coalescence}. 

The structural diversity index of the network $G$ is defined as the ratio between its expected meeting time $\langle M_G \rangle$ and its number of vertices, or size, $\vert V(G) \vert$:
\begin{equation}
    \Delta(G) = \frac{\langle M_G \rangle}{\vert V(G) \vert} \ .
\end{equation}
On the surface, $\Delta(G)$ is a scale invariant measure of the ease of meeting during a random walk on the network. 
Although this metric is solely based on the network's structure, it offers powerful predictions of the network's propensity to support socio-diversity.

The fascinating relationship between the structural diversity index $\Delta(G)$ and socio-diversity is best understood through a simple model of cultural evolution, commonly known as the voter model \cite{liggett1985voter}. As before, we have a population occupying the vertices of a network $G$ with edges symbolizing various undirected and reciprocal relationships, such as friendships, acquaintances, or interactions. 

At the beginning, each individual $i$ of the population displays a distinct meme $m_i$. 
A meme symbolizes an individual cultural trait that can vary---such as political beliefs or musical tastes.
Cultural evolution is assumed to happen at discrete time steps. During each step, individuals simultaneously modify their meme in one of two ways:
\begin{enumerate}
    \item[(i)] by imitating the current meme of a randomly chosen neighbor, which occurs with a probability of $1-r$.
    \item[(ii)] by inventing a novel meme that has not been previously existed, with a probability of $r$.\footnote{Innovation need not imply a groundbreaking cultural shift but might just be a nuanced variation of an existing meme.}
\end{enumerate}
The parameter $r$, with $0\le r \le 1$, is referred to as the innovation rate. 
It measures the equilibrium between the two key evolutionary forces of imitation and innovation. 
Higher values of $r$ stimulate innovation, while lower values strengthen imitation.

Figure \ref{figure:illustration_vm}D-F illustrates cultural evolution in a small social network over three time steps $t=1,2,3$. 
In Panel D, each individual displays a distinct meme, represented by the color of their respective vertex. We focus on the journey of a specific individual, Elon. 
Transitioning from $t=1$ (Panel D) to $t=2$ (Panel E), we observe that Frank and Carl imitate Elon's blue meme (indicated by blue arrows). 
In contrast, Elon innovates, introducing a new meme: the black meme. 
Globally, the meme landscape has undergone a transformation. 
Elon's black meme enters the scene, the green and yellow disappear, and the blue meme gets increased attention. Advancing from $t=2$ (Panel E) to $t=3$ (Panel F), Elon opts to imitate Frank by embracing the blue meme. Concurrently, Frank, Carl, and Darcy find Elon's new meme appealing and imitate it. 
As a result, only two dominant memes emerge: blue and black.

In short, as cultural evolution unfolds over time, individuals either innovate (as Elon) by creating new memes or imitate (as everyone else) by adopting existing memes from their peers. 
These processes continually reshape the meme landscape in the population, influencing the population's overall socio-diversity. 
For instance, it is evident that the population in Figure \ref{figure:illustration_vm}'s Panel D is inherently more diverse than that in Panel F. 
Yet, to quantify this difference in diversity, one requires a specific measurement of socio-diversity.

In this study, we adopt a well-known diversity measure in ecology: Simpson's diversity index \cite{simpson1949measurementa}. 
This index takes into account both the number of memes, as well as their relative abundance. 
It is defined as the probability that two randomly selected individuals in the population display different memes. 
The value ranges from $0$ to $1$. 
When the population's memes are homogeneous, the probability that two randomly selected individuals exhibit different memes is small. 
Consequently, Simpson's diversity index is close to 0. Conversely, when the population's memes are diverse, the probability that two randomly selected individuals display different memes is large. 
Accordingly, Simpson's diversity index is close to $1$. 

In the context of our cultural evolution model, Simpson's diversity index at fixed time $t$ can be computed as:
\begin{equation}
    D(t) = 1 - \sum_{m} p_m(t)^2 \ . 
\end{equation}
Herein, the summation is over all memes present at time $t$, and $p_m(t)$ is the fraction of individuals who display meme $m$ at that time. 
For illustration, the socio-diversity $D(t=1)$ of the (highly diverse) population portrayed in Figure \ref{figure:illustration_vm}D is $D(1) = 0.833$, while that of the (more homogeneous) population in Figure \ref{figure:illustration_vm}F is $D(3) = 0.5$. 
In summary, higher values of $D(t)$ reflect greater socio-diversity within the population.

Returning to our fundamental research question, we can now frame it more precisely: how does the network structure $G$ of a population influence its socio-diversity $D(t)$? 
Our answer is the following simple and elegant mathematical equation, which is a generalization of results by Aldous and collaborators \cite{aldous2013probability, liggett1985voter}. This equation links the meeting time in the random social exploration game (depicted in Figure \ref{figure:illustration_vm}A-C) to the long-term socio-diversity in our cultural evolution model (as shown in Figure \ref{figure:illustration_vm}D-F). The derivation of this equation arguably represents the most intricate part of our analysis and can be found in Methods:
\begin{equation}
    \label{eq:duality_short}
    D_{\infty} \defeq \lim_{t\rightarrow \infty} \frac{1}{t} \sum_{s \leq t} D(s) \approx 1 -  e^{-2 \alpha \Delta(G)} \ .
\end{equation}
The left-hand side of the equation introduces $D_{\infty}$, the average population wide socio-diversity over an extended period of time (henceforth the population's \emph{expected socio-diversity}). 
This quantity is the focal point of our exploration. 
We aim to understand what elements of a population's network structure affect its expected socio-diversity $D_{\infty}$.

\eqref{eq:duality_short} reveals that the expected socio-diversity is determined by two fundamental factors: first, the per-capital innovation rate $\alpha = r\vert V(G) \vert$. 
Unsurprisingly, a higher per-capita innovation rate leads to greater socio-diversity; second, and most importantly, the network structure $G$, as captured by the structural diversity index $\Delta(G)$. 
When $\Delta(G)$ is high, the expected socio-diversity tends to be high as well, nearing its maximum of $1$. 
Conversely, when $\Delta(G)$ is low, the expected socio-diversity is low, actually close to the minimum value of 0. 
This codependence relationship suggests that the structural diversity index $\Delta(G)$ captures the connection between network structure, upon which it depends, and socio-diversity, which it influences.

\begin{figure*}
    \centering
    \includegraphics[width=\textwidth]{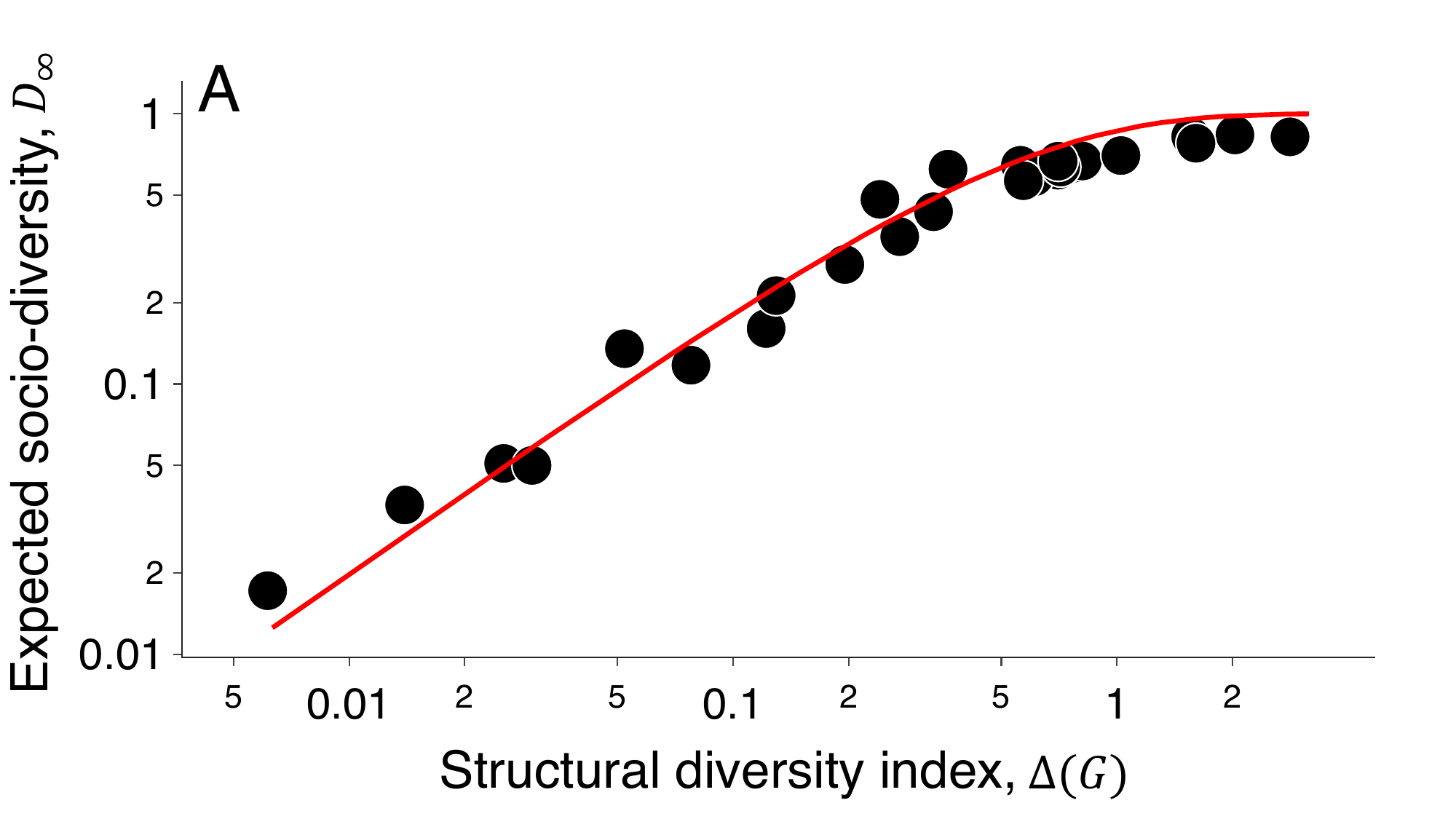}
    \caption{Correlation of expected socio-diversity $D_{\infty}$ with the structural diversity index $\Delta(G)$. We simulated the cultural evolution model across various real-world social networks $G$ to determine $D_{\infty}$. Each dot represents a simulation for a distinct social network. The red line depicts the curve $1 - e^{-2\Delta(G)}$, our theoretical estimate for $D_{\infty}$ derived from \eqref{eq:duality_short} using $\alpha = 1$ (or $r = 1/\lvert V(G) \rvert)$. Remarkably, the structural diversity index \emph{predicts} expected socio-diversity levels quite accurately, as evidenced by observations scattering around the red line. See Methods for simulation parameters and descriptions of the social networks.}
    \label{fig:duality}
\end{figure*}

Figure \ref{fig:duality} plots the relationship between expected socio-diversity $D_{\infty}$ and the structural diversity index $\Delta(G)$ for various real-world social networks $G$. 
In a log-log plot, a clear saturating relationship is found, which aligns with our analytical prediction based on \eqref{eq:duality_short} (see red line). 

\subsection*{Amplifiers and suppressors of socio-diversity}
\eqref{eq:duality_short} captures the complex interplay between socio-diversity and network structure by expressing the expected socio-diversity $D_{\infty}$ as a function of a quantity that only depends on network structure, namely, the structural diversity index $\Delta(G)$. 
By increasing the structural diversity index $\Delta(G)$ in \eqref{eq:duality_short}, we observe an increase in expected socio-diversity $D_{\infty}$. 
Hence, networks with large structural diversity index tend to favor socio-diversity, whereas those with a small one tend to obstruct it. 

But what should be considered ``large'' or ``small'' structural diversity indices? 
Large and small are typically defined with respect to a benchmark. 
The natural benchmark here is the complete network $K$. 
The complete network reflects the total absence of social structure. 
There are no communities, no clusters, and no differences between individual social positions; the population is structurally homogeneous. 
Comparing the structural diversity index of an arbitrary network $G$ with that of an equally sized complete network $K$ informs us about how the network structure affects the index, and, consequently, socio-diversity.  

The structural diversity index of the complete network satisfies $\Delta(K) = 1$, independent of its size (see Methods for an explanation). 
If, for a network structure $G$, we have $\Delta(G) <\Delta(K) = 1$, \eqref{eq:duality_short} suggests that the population's expected socio-diversity is lower than if the population were unstructured. 
In other words, all else equal, the variety of memes in a population with structure $G$ is expected to be lower than in a population with no structure. 
Hence, networks $G$ with $\Delta(G) < 1$ can be said to (structurally) \emph{suppress socio-diversity} (see Ref. \cite{lieberman2005evolutionary}). 
In contrast, networks $G$ with $\Delta(G) > \Delta(K) = 1$ can be said to (structurally) \emph{amplify socio-diversity}. Indeed, according to \eqref{eq:duality_short}, the population's expected socio-diversity is higher than if the population were unstructured. 

\begin{figure}[t]
    \centering
    \includegraphics[width=\linewidth]{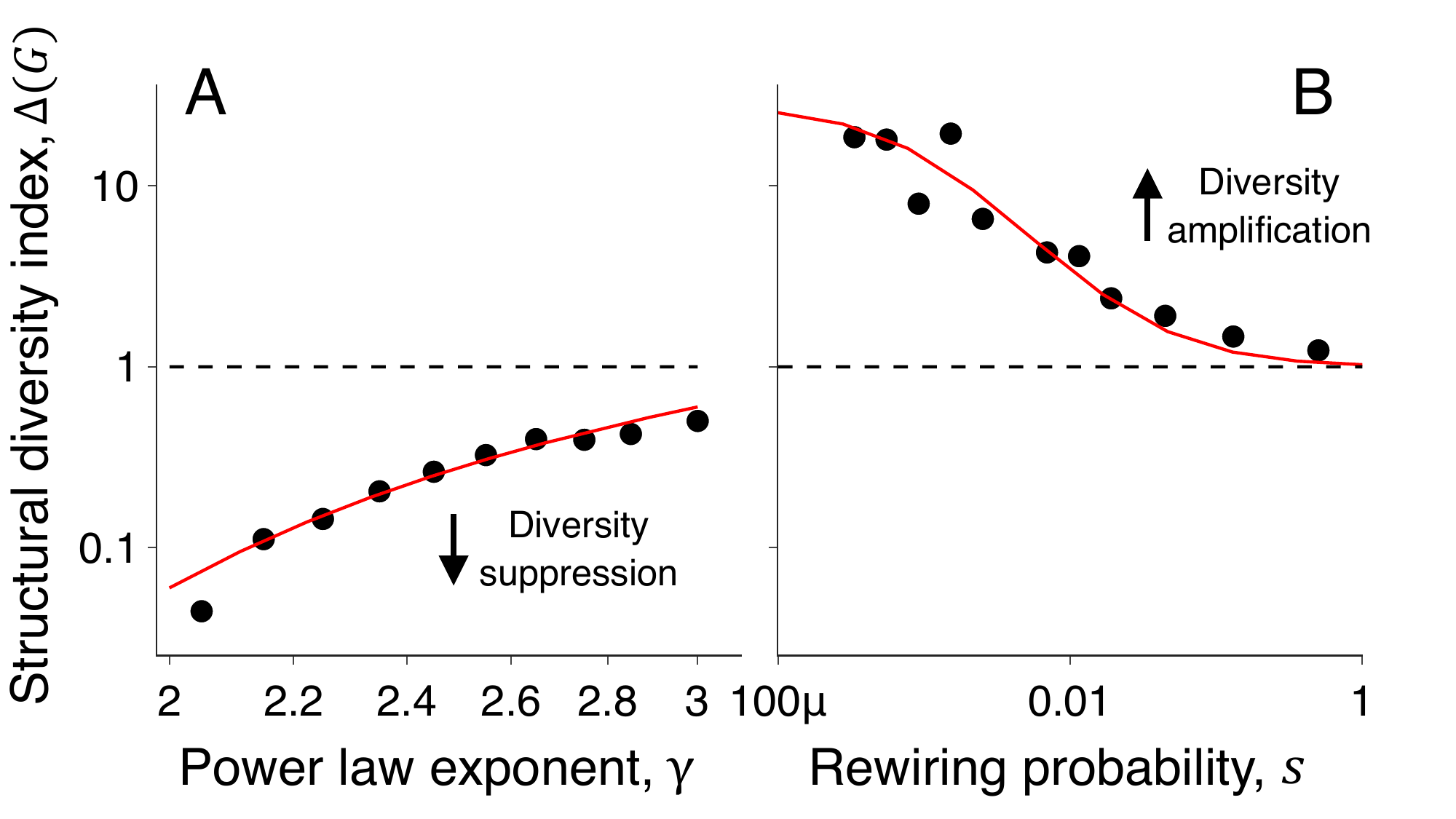}
    \caption{Numerical simulation (dots) and analytical estimates (red lines) of the structural diversity index of (A) scale-free and (B) Watts-Strogatz networks. These are plotted against (A) the power-law exponent $\gamma$ and (B) the rewiring probability $s$. In (A), the red lines show the equation $\Delta(G_{\gamma}) = b \cdot \lvert V(G_{\gamma}) \rvert^{-a \cdot (3-\gamma)/(\gamma-1)}$ where $a$ and $b$ are obtained by ordinary least squares fit. In (B), the red line represents the approximation in \eqref{eq:deltag_ws}. Scale-free networks tend to suppress socio-diversity ($\Delta(G_{\gamma}) < 1$). Specifically, greater heterogeneity in the degree distribution (i.e., a smaller exponent $\gamma$) induces greater suppression of socio-diversity (i.e., smaller values of $\Delta(G_{\gamma})$). Conversely, Watts-Strogatz networks tend to amplify socio-diversity ($\Delta(W_s) > 1$). However, socio-diversity amplification is reduced as more long-range connections are established or/and more randomness is inserted (i.e., as the rewiring probability $s$ increases). Supplementary Movie 1 offers a visual comparison of cultural evolution on scale-free and Watts-Strogatz networks (see Section F.1 of the SI for the movie's caption). See Methods for simulation parameters.}
    \label{fig:SFWS}
\end{figure}

Scale-free networks $G_{\gamma}$ are networks characterized by a power-law degree distribution $P(k) \sim k^{-\gamma}$. 
In the Methods, we show that the structural diversity index of scale-free networks with exponent $\gamma$, with $2 \leq \gamma \leq 3$, satisfies
\begin{equation}
    \label{eq:deltag_sf}
    \Delta(G_{\gamma}) \leq \lvert V(G_{\gamma})\rvert^{-\frac{3-\gamma}{\gamma-1}} \leq 1 \ .
\end{equation}
Therefore, scale-free networks tend to suppress socio-diversity. 
Moreover, as illustrated in Figure \ref{fig:SFWS}A, diversity suppression intensifies as the scale-free network becomes more degree-heterogeneous (i.e., as the exponent $\gamma$ decreases).

Watts-Strogatz networks $W_s$ interpolate between regular lattices and random networks by means of a parameter $s \in [0,1]$, called rewiring probability \cite{watts1998collective}. 
The structural diversity index of these networks is roughly
\begin{equation}
    \label{eq:deltag_ws}
    \Delta(W_s) \approx \frac{s + 1/\langle k \rangle^2}{s + 1/\lvert V(W_s) \rvert} \geq 1 \ ,
\end{equation}
where $\langle k \rangle$ denotes the network's average degree  (see Methods). 
Hence, Watts-Strogatz networks tend to amplify socio-diversity, and this amplification weakens as randomness increases (i.e., as the rewiring probability $s$ becomes larger)--- see Figure \ref{fig:SFWS}B.

\subsubsection*{Characteristics of real-world networks that amplify and suppress socio-diversity}
\begin{figure*}[htp]
    \centering
    \includegraphics[width=\textwidth]{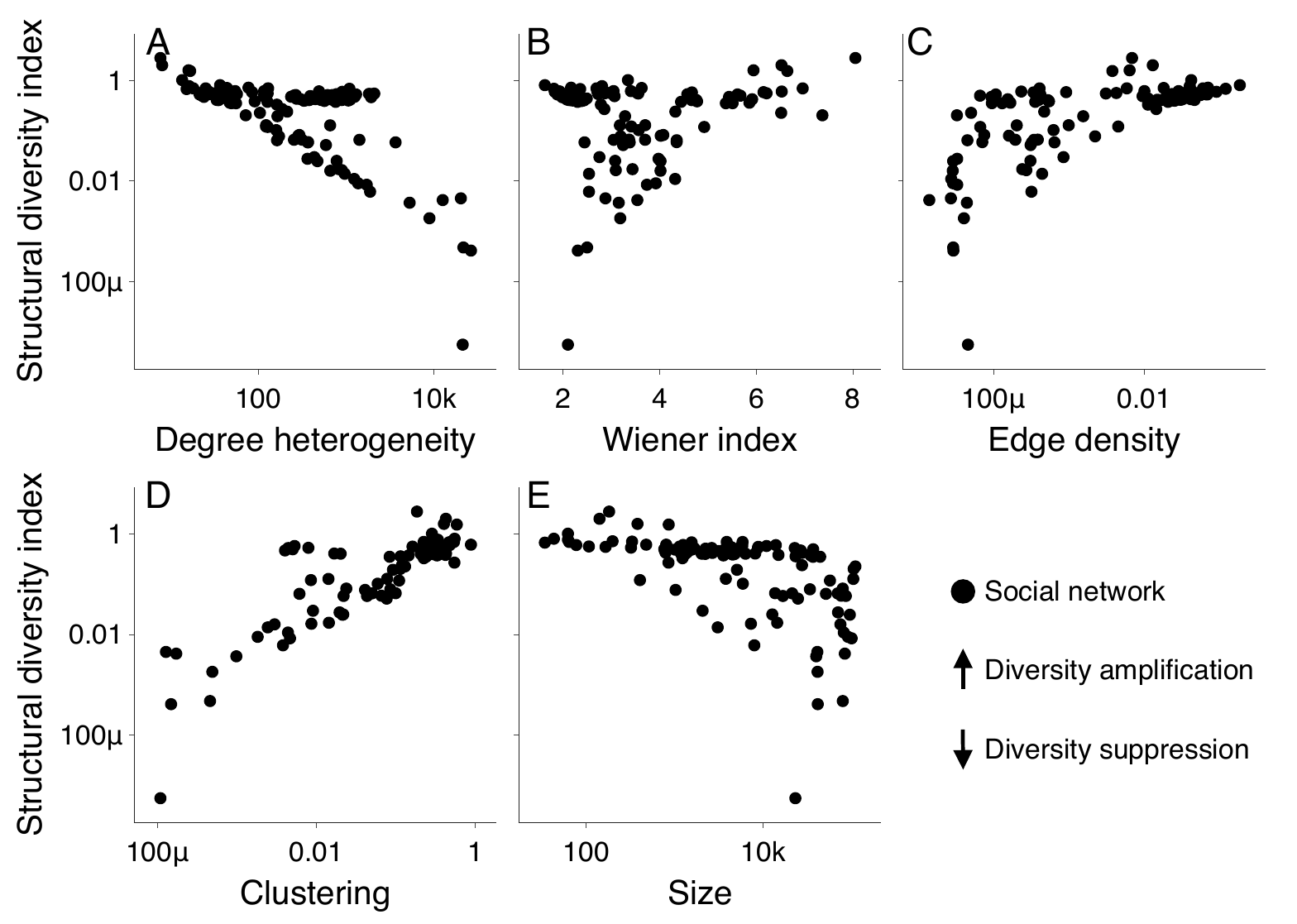}
    \caption{Relationship between the structural diversity index and (A) the degree-heterogeneity $\kappa(G)$, (B) the Wiener index $W(G)$, (C) the edge density $e(G)$, (D) the clustering coefficient $c(G)$, and (E) the size $\lvert V(G) \rvert$ of a network $G$ (see main text for definitions). Analyzing a variety of social networks we find that high degree-heterogeneity tends to suppress socio-diversity. In contrast, high clustering, Wiener index and edge density tend to amplify it. The effect of network size is more intricate (see text for details). See Methods for descriptions of the networks.}
    \label{fig:gprops}
\end{figure*}

Let us broaden the scope of our analysis from the previous examples and ask: what general characteristics of networks amplify or suppress expected socio-diversity? 
Figures \ref{fig:gprops}A-E plot the structural diversity index $\Delta(G)$ against five well-known properties of social networks---degree-heterogeneity, Wiener index, edge density, clustering, and size---for a wide range of real-world social networks $G$. 
Table \ref{table:regression} presents the outcomes of five regression models. These models help quantify the correlations shown in Figure \ref{fig:gprops} and evaluate their robustness.

Figure \ref{fig:gprops}A reveals a strong negative correlation between the structural diversity index $\Delta(G)$ and degree-heterogeneity $\kappa(G)$, measured as the ratio of the second and first moments of $G$'s degree distribution \cite{barrat2008dynamical}. 
As degree heterogeneity increases, the structural diversity index decreases. This correlation is quite robust. 
It persists even after accounting for other network properties (see Table \ref{table:regression}) and evaluating alternative measures of inequality in the distribution of connection, such as the Gini Index (see Section B of the SI). 
At its core, this suggests that high degree-heterogeneity tends to suppress socio-diversity. This phenomenon has an intuitive explanation: large-degree vertices (``VIPs'', ``hubs'', ``influencers'', or ``hyperinfluentials'' \cite{watts2007influentials}) are crucial---either as initiators or early adopters---in triggering large imitation cascades \cite{watts2007influentials}. 
This eventually ends up reducing socio-diversity.

Figure \ref{fig:gprops}B portrays a positive correlation between the structural diversity index $\Delta(G)$ and the Wiener index $W(G)$. 
Specifically, when the Wiener index has high values, the structural diversity index also tends to be high.
The regressions in Table \ref{table:regression} confirm this observation. Moreover, Model 5 in this same table reveal an increase of this correlation when controlling for the effects of other network attributes. 
From a wider viewpoint, these findings indicate that large network distances between individuals tend to amplify socio-diversity. 
The reason is intuitive: large distances obstruct meme spreading, a phenomeon testified by the geographical clustering of most cultural forms.

Figure \ref{fig:gprops}C and Model 3 in Table \ref{table:regression} reveal a positive relationship between the structural diversity index and edge density $e(G)$, defined as the proportion of existing edges to potential edges within the network. 
Table \ref{table:regression} (Model 5) demonstrates that this correlation intensifies when accounting for other network characteristics, suggesting that edge density may play an important role in the regulation of socio-diversity. 
Broadly, in a similar vein to how it fosters meritocracy \cite{borondo2014each}, edge density appears to encourage socio-diversity. 
These findings align with theories arguing that a greater number of connections reduces the influence of each individual connection, thus diminishing the likelihood of large imitation cascades \cite{granovetter1978threshold}.

\begin{table}\centering
    \caption{\textbf{Dependent variable: Log(structural diversity index)}. The regression models elucidate the correlations between the structural diversity index and network characteristics as illustrated in Figure \ref{fig:gprops}: (i) the correlations with degree-heterogeneity, the Wiener index, and edge density are robust. (ii) The correlation with clustering fades when accounting for degree-heterogeneity, Wiener index, size and edge density. (iii) The correlation with size reverses when factoring in other network characteristics. All variables are standardized for a direct comparison between regression coefficients. See main text for interpretations of these results and Methods for technical details about the regressions.}
    \begin{tabular}{@{}lccccc@{}}
    \toprule
    \textbf{Variable} & \textbf{Model 1} & \textbf{Model 2} & \textbf{Model 3} & \textbf{Model 4} & \textbf{Model 5}\\ 
    \midrule
    Log(degree heterogeneity) & -0.67*** & & & & -0.56*** \\
                       & (0.07) & & & & (0.06)   \\
    Log(wiener index)  & & 0.03 & & & 0.37*** \\
                        & & (0.09) & & & (0.08) \\
    Log(edge density)  & & & 0.65*** & & 1.38***\\
                       & & & (0.07) & & (0.12) \\
    Log(clustering)    & & & & 0.84*** & 0.02 \\
                       & & & & (0.05)  & (0.06)  \\
    Log(size)  & & & & & 0.70*** \\
                       & & & & & (0.07) \\
    \midrule
    $\#$ observations & 119 & 119 & 119 & 119 & 119   \\          
    $R^{2}$           & 0.45  & 0.0  & 0.43  &  0.72 & 0.93   \\
    Adjusted $R^{2}$  & 0.44  & 0.0  & 0.42  &  0.71 & 0.92   \\
    \bottomrule
    \addlinespace[1ex]
    \multicolumn{3}{l}{\textsuperscript{***}$p<0.001$, 
      \textsuperscript{**}$p<0.01$, \textsuperscript{*}$p<0.1$}
    \end{tabular}
    \label{table:regression}
    \end{table}

Figure \ref{fig:gprops}D shows a positive correlation between the structural diversity index $\Delta(G)$ and clustering, as measured by the clustering coefficient $c(G)$ \cite{watts1998collective}. 
On the surface, higher levels of clustering, imply a larger structural diversity index. However, Model 5 of Table \ref{table:regression} highlights that this correlation fades when factoring in other network characteristics. 
This suggests that these network characteristics might mediate the amplifying effect of clustering on socio-diversity. 
To break it down, a large network with high average inter-node distances and significant edge density can support socio-diversity irrespective of its clustering levels. 
However, in our dataset, most dense networks exhibit high clustering. 
Consequently, it is complicated to disregard the significance of clustering entirely. 
This viewpoint is further supported by a straightforward mechanism connecting clustering with socio-diversity: clusters, when they are meme-homogeneous, obstruct consensus formation by increasing the persistence of individual memes and decreasing the exposure to new memes. 

Finally, Figure \ref{fig:gprops}E highlights a negative correlation between the structural diversity index and network size, suggesting that large networks might suppress socio-diversity. 
However, a closer look (see Table \ref{table:regression}, Model 5), reveals a shift to a positive correlation when factoring in all the discussed network characteristics. 
This sheds some light on the nuanced relationship between the structural diversity index and network size: Size intrinsically boosts the index, perhaps due to factors like increased overall innovation in larger networks. 
Yet, as networks grow, they become more sparse because maintaining connections is costly. 
And this decrease in edge density is likely responsible for the initial negative correlation observed in Figure \ref{fig:gprops}E.

\section*{Discussion}
Understanding the interplay between network structure and socio-diversity is crucial, as the latter has numerous positive and negative implications for society. 
In this article, we have made some steps to understand this. 
We have found that: (i) A simple index, the structural diversity index, captures the complex interplay between network structure and socio-diversity. 
(ii) Network characteristics can amplify or suppress socio-diversity: high degree-heterogeneity, as in scale-free networks, tends to suppress it, while high local clustering, large inter-node distances, and significant edge density tend to amplify it. 
For clarity, we explored the voter model, one of the simplest models of cultural evolution. However, in Section C of the SI, we show that qualitatively similar results hold for other fundamental models such as Axelrod's model \cite{axelrod1997disseminationa}, Sznajd's model \cite{sznajd-weron2000opinion} and the (discrete) bounded confidence model \cite{hegselmann2005opinion, lorenz2007continuous} (see also the review in Ref. \cite{castellano2009statisticala}). 

\begin{figure}[htp]
\includegraphics[width=\linewidth]{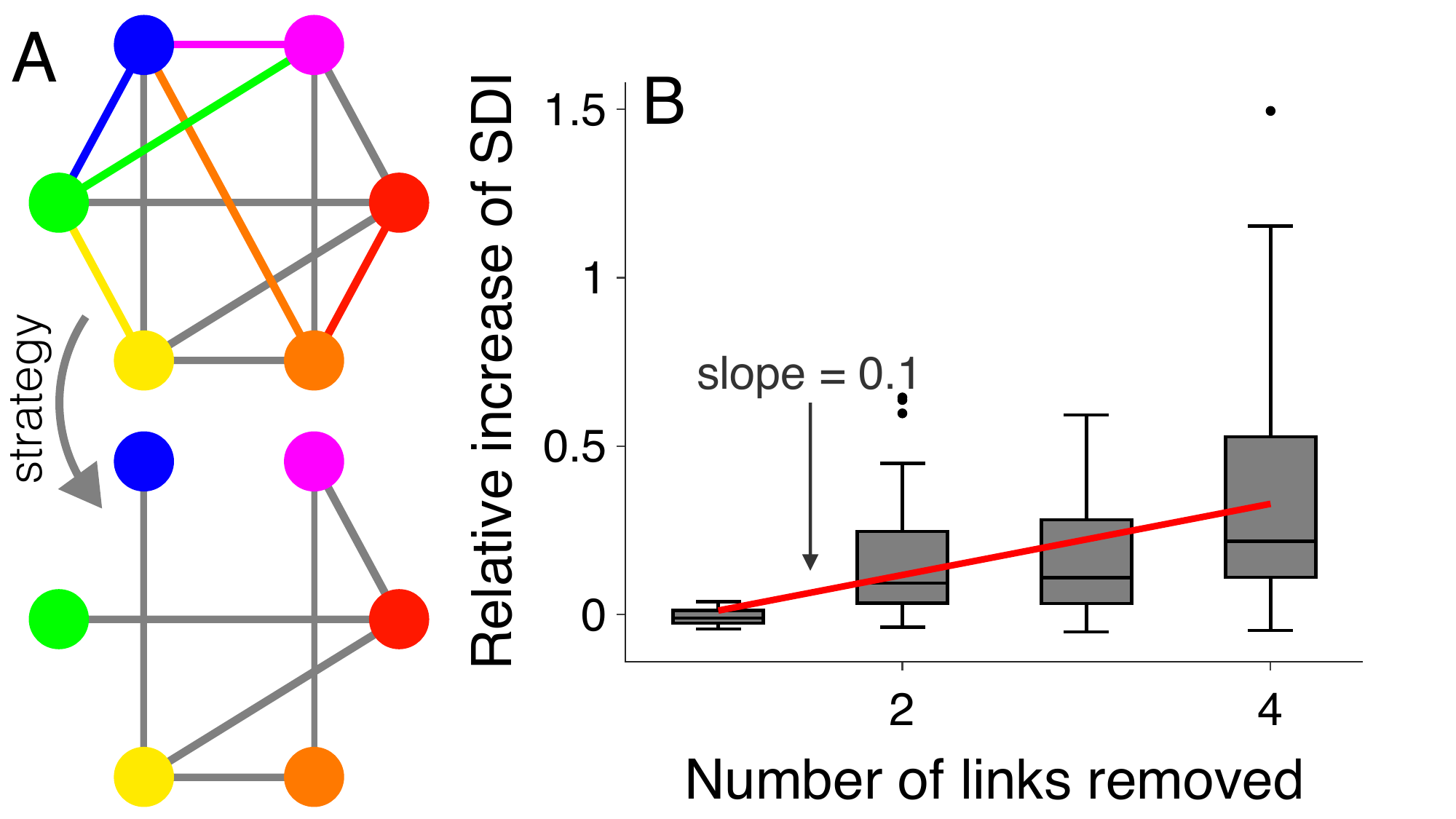}
\caption{Removing links to highly connected individuals may increase the structural diversity index. \textbf{(A)} Illustration of a simple strategy to raise the structural diversity index: each individual in the network $G$ (top) removes links to her $h = 1$ most connected neighbors (tie sorting is random), resulting in the network $G_h$ (bottom). \textbf{(B)} Percentage change $(\Delta(G_h)-\Delta(G))/\Delta(G)$ in the structural diversity index after applying the procedure outlined in (A) to a network $G$. This simple procedure leads to remarkable increases of the structural diversity index, even when the number of removed connections per individual is small. On average, increasing $h$ by one leads to a $10\%$ increase in the index. See Methods for details about network data, box plots and regression values.}
\label{fig:edge_removal}
\end{figure}

Our work suggests numerous future research directions (see also Section E of the SI). 
First, an understanding of the consequences that specific characteristics of networks have for socio-diversity implies opportunities for change. 
For example, our results hint towards possible ways of transforming social networks to sustain greater socio-diversity. 
For instance, when an increase in degree-heterogeneity causes a reduction in socio-diversity, a simple, decentralized strategy such as ``stop following the $h$ most connected VIPs in your social network channel'' (or, for short, ``don't follow leaders'' \cite{livan2019don}) can be surprisingly effective in sustaining it, as shown in Figure \ref{fig:edge_removal}. 
Future research may explore further kinds of network modification strategies that amplify or suppress socio-diversity. For example, how can one leverage the fact that clustering amplifies socio-diversity? 

Second, and most importantly, our findings are primarily rooted in models of cultural evolution, rather than in actual experimental data. 
It is, therefore, essential to remember that all models inherently simplify the complexities of human interactions. 
As such, empirical data may reveal nuances that go beyond the narratives presented in our study. 
Thus, we encourage follow-up research to validate our conclusions through lab or online experiments. 
In essence, a pivotal question lingers: Can the insights about the relationship between network structure and socio-diversity gained from our modeling and simulations be replicated in an experimental environment? 

A third key direction of future research is to deepen our understanding of what levels of socio-diversity are beneficial for distinct social systems. 
Our work illustrates how one can change social networks to promote or reduce socio-diversity, but it does not tackle whether more or less diversity is desirable. Socio-diversity can bring both advantages and challenges: while an overabundance might lead to division and conflict, too little could hinder innovation and collective intelligence. 
Pinpointing the ideal balance is complex, necessitating a careful consideration of socio-diversity's multifaceted effects. 
Nevertheless, given its profound implications for societal dynamics, it is high time that we appreciate the importance of socio-diversity and explore new ways of shaping it.

\section*{Methods}
\subsection*{Proof of \eqref{eq:duality_short}}
We will present a direct proof of this \eqref{eq:duality_short}, based on the fundamentals of random walk theory.

The proof interlinks the model of cultural evolution discussed above with the concept of an $r$-random walk on a graph $G$.
This $r$-random walk closely resembles a conventional random walk, but with an added twist: at every step, there is a probability $r$ of the walk ``halting''. 
For a more vivid picture, imagine, as Karl Pearson and Lord Rayleigh \cite{pearson1905problem, rayleigh1905problem}, a drunkard wandering through an urban street network. 
At every intersection, he randomly selects a street and heads towards the next crossing. 
The nuance in the $r$-random walk lies here: on any given street, the drunkard might come across a bar he fancies with probability $r$, leading him to leave the street network indefinitely.

The core relationship between $r$-random walks and the cultural evolution model discussed above is summarized in the following equation. 
This equation is based on the principle of ``voter model duality'' \cite{liggett1985voter}. It is a straightforward generalization of Aldous' findings for traditional random walks \cite{aldous2013interacting}:
\begin{equation}
\label{eq:duality_div_mp}
    1 - D(t) = p_r(t) \ .
\end{equation}
In this context, $1-D(t)$ represents the likelihood that, at time step $t$ in the cultural evolution model, a pair of randomly chosen individuals both exhibit the same meme. 
On the other hand, $p_r(t)$ indicates the probability that two $r$-random walks, started with uniform probability across the vertices of $G$, meet before completing $t$ steps.

\eqref{eq:duality_div_mp} leaves us with the task of understanding $p_r(t)$. 
Since it simplifies the argument and since our main interest concerns the large-time behavior of the system, we focus on $p_r(\infty) = \lim_{t \rightarrow \infty} p_r(t)$. 
This effectively means that we are exploring the probability that two $r$-random walks on the graph $G$ meet before either one halts.

In mathematical symbols, ``the probability that two $r$-random walks on the graph $G$ meet before either one halts'' can be expressed as:
\begin{equation}
    \label{app:eq:duality_pm_meeting}
    p_r(\infty) = P(M_G < \min(S_1, S_2)) \ .
\end{equation}
In this equation, $M_G$ is the meeting time of the graph $G$. 
This quantity has been defined as the number of steps before two uniformly started \emph{traditional} random walks on $G$ visit the same vertex simultaneously. 
Meanwhile, $S_1$ and $S_2$ are geometric random variables with a success probability $r$. 
These random variables count the number of steps taken by the first and second random walks, respectively, before they come to a halt.

To derive \eqref{eq:duality_short}, we approximate:
\begin{equation*}
\label{app:eq:approximation}
 P\big(M_G < \min(S_1, S_2)\big) \approx P\big(\langle M_G \rangle < \min(S_1, S_2)\big) \ .
\end{equation*}
Section A of the SI discusses this approximation in detail.
Next, since $S_1$ and $S_2$ are geometrically distributed random variables with success probability $r$, $\min(S_1,S_2)$ is a geometrically distributed random variable with success probability $q(r) = 2r-r^2$. Therefore,
\begin{align*}
     P\big(\langle M_G \rangle < \min(S_1, S_2)\big) &= (1-q(r))^{\langle M_G \rangle} \approx e^{-q(r)\langle M_G \rangle} \ .
\end{align*}
When $r \ll 1$, $q(r) \approx 2r$. Hence, $e^{-q(r)\langle M_G \rangle} \approx e^{-2r\langle M_G \rangle}$. 
The hypothesis $r \ll 1$ is convenient for presentation because it makes interpretation more straightforward. 
However, it is not necessary for the paper's conclusions to hold. In fact, replacing $2r$ by $q(r) = 2r - r^2$ yields very similar results. 

Drawing upon our prior arguments, we have established:
\begin{equation}
\label{eq:pr_infty}
    p_r(\infty) = P\big(M_G < \min(S_1, S_2)\big) \approx  e^{-2r\langle M_G \rangle} \ .
\end{equation}
Now, \eqref{eq:duality_div_mp} shows that $1 - D(t) = p_r(t)$. Therefore, 
\begin{equation}
    p_r(\infty) = \lim_{t\rightarrow \infty} p_r(t) = \lim_{t\rightarrow \infty} 1 - D(t) =  1 - \lim_{t \rightarrow \infty} \sum_{s \leq t} D(s) = 1 - D_{\infty} \ .
\end{equation}
From this relationship and \eqref{eq:pr_infty}, it is clear that:
\begin{equation}
D_{\infty} = 1 - p_r(\infty) = 1 - e^{-2r\langle M_G \rangle} \ .
\end{equation}
Finally, replacing the innovation rate $r$ with the per-capita innovation rate $\alpha = r \vert V(G) \vert$ and recalling that the structural diversity index is defined as the ratio $\Delta(G) = \langle M_G \rangle / \vert V(G) \vert$ we obtain the desired equation
\begin{equation}
D_{\infty} = 1 - e^{-2\alpha \Delta(G) } \ .
\end{equation}
\subsection*{Structural diversity index of complete, scale-free, and Watts-Strogatz networks}
For a network $G$, we defined the structural diversity index by $\Delta(G) = \langle M_G \rangle/ \lvert V(G) \rvert$, where $M_G$ is the meeting time of two (uniformly started) random walks on $G$ and $\lvert V(G) \rvert$ is $G$'s vertex count. 
Let us discuss estimates for the structural diversity index of complete, scale-free, and Watts-Strogatz networks. 

\bmhead{Complete networks} For the complete network $K$ with $\lvert V(K) \rvert$ vertices, the structural diversity index can be computed exactly. 
In each step, two random walks on $K$ have a probability $1/\lvert V(K) \rvert$ of moving to the same vertex and, hence, of meeting. 
Consequently, the meeting time $M_K$ of the walks is geometrically distributed with success probability $1/\lvert V(K) \rvert$. In particular, $\langle M_K \rangle = \lvert V(K) \rvert$.

\bmhead{General bounds}For a general network $G$, exact analytical expressions for the meeting time $M_G$ are not in reach. However, good upper and lower bounds are available. 

On the one hand, Cooper et al. \cite{cooper2013coalescing} provide an upper bound for the average meeting time $\langle M_G \rangle$:
\begin{equation}
\label{methods:eq:upperbound_mt}
    \langle M_G \rangle = O \Big(\frac{1}{1-\lambda_2} \big( 2\log(\lvert V(G) \rvert) + \lvert V(G) \rvert\frac{\langle k \rangle^2 }{\langle k^2 \rangle} \big) \Big) \, .
\end{equation}
Herein, $O$ is the standard asymptotic notation for ``is asymptotically dominated by''; $\langle k \rangle$ and $\langle k^2 \rangle$ denote respectively the first and second moments of $G$'s degree distribution; and $\lambda_2$ indicates the second largest eigenvalue of the transition matrix $P$ of the random walk on $G$ (i.e., the matrix $P(i,j) = 1/\deg(i)$ for all vertices $i, j$ of $G$).

On the other hand, Aldous \cite{aldous1991meeting} provides a lower bound for the average meeting time $\langle M_G \rangle$:
\begin{equation}
\label{methods:eq:lowerbound_mt}
    \langle M_G \rangle = \Omega\Big(\frac{\lvert E(G) \rvert}{D_{\max}} \Big)
\end{equation}
Herein, $\Omega$ is the standard asymptotic notation for ``asymptotically dominates''; $D_{\max}$ is the maximum degree in the graph $G$; and $\lvert E(G) \rvert$ is the number of edges.

\bmhead{Scale-free networks}
For a scale-free network $G_{\gamma}$ with exponent $\gamma$, different bounds for $1/(1-\lambda_2)$ exist, depending on the model with which the network is constructed \cite{gkantsidis2003conductance, mihail2003certain}. However, such bounds are typically polynomials in $\log(\lvert V(G_{\gamma})\rvert )$. 
Therefore, neglecting logarithmic terms, \eqref{methods:eq:upperbound_mt} suggests that $\langle M_{G_{\gamma}} \rangle = O(\lvert V(G_{\gamma}\rvert ) \langle k \rangle ^2 /\langle k^2 \rangle)$. Hence, for large networks $G_{\gamma}$ we have
\begin{equation}\label{app:eq:threshold_sf}
    \Delta(G_{\gamma}) \leq \frac{\langle k^2 \rangle }{\langle k \rangle^2}\ .
\end{equation}
Herein, $\langle k\rangle$ and $\langle k^2 \rangle$ denote the first and second moments of $G_{\gamma}$'s degree distribution. 
Note that $\langle k \rangle$ and $\langle k^2 \rangle$ depend on the exponent $\gamma$. We have calculated their values in terms of $\lvert V(G_{\gamma}) \rvert$ and $\gamma$ following the guidelines provided in Chapter 6.4 of Ref. \cite{barrat2008dynamical}. 
This yields the bound for $\Delta(G_{\gamma})$ reported in \eqref{eq:deltag_sf}:
\begin{equation}
      \Delta(G_{\gamma}) \leq
      \begin{cases} 
      1 & \mbox{if }\gamma \geq 3 \\
      \lvert V(G_{\gamma}) \rvert^{-\frac{3-\gamma}{\gamma-1}} & \mbox{if } 2 < \gamma < 3 \\
    \lvert V(G_{\gamma}) \rvert^{-1} & \mbox{if } \gamma \leq 2 
    \end{cases} \ .
\end{equation}

\bmhead{Watts-Strogatz networks}
\label{methods:sec:ws-sdi}
A Watts-Strogatz network $W_s$ with rewiring probability $s$ and average degree $\langle k \rangle$ has $\lvert E(W_s) \rvert = \langle k \rangle \lvert V(W_s) \rvert$ edges and maximum degree $D_{\max} \approx \langle k \rangle$. 
Therefore, according to \eqref{methods:eq:lowerbound_mt}, the average meeting time satisfies $\langle M_{W_s} \rangle = \Omega(\lvert V(W_s) \rvert)$. 
Consequently, for large networks $W_s$, we retrieve the lower bound reported in \eqref{eq:deltag_ws}:

\begin{equation}
    \Delta(W_s) \geq 1 \ .
\end{equation}

The following explicit formula captures the relationship between the structural diversity index and the rewiring probability $s$ fairly well:
\begin{equation}
    \Delta(W_s) \approx \frac{s + 1/\langle k \rangle^2}{s + 1/\lvert V(W_s) \rvert} \ .
\end{equation}
However, we could not find a theoretical derivation of this approximate relationship.

\subsection*{Network data}
All networks have been handled using graph-tool \cite{peixoto2014graphtool} or networkx \cite{hagberg2008exploring}. 
The network data employed in this work are described in Section B of the SI and are freely available at \url{https://networks.skewed.de}. 
A Python package enabling the fast numerical computation of the structural diversity index developed by the first author is available through PyPi \url{https://pypi.org/project/structural-diversity-index/}. 
See Section D of the SI for more information. 

\subsection*{Parameters and Specifications for Figures}
\bmhead{Figure \ref{fig:duality}} For selected real-world networks $G$ from our network dataset (see Table B.1 and Table B.2 of the SI for details), we computed the expected socio-diversity $D_{\infty}$ by simulating the cultural evolution model with parameter $r = 1/\lvert V(G) \rvert$ for $10 \cdot \lvert V(G) \rvert$ steps 20 times; 
(ii) the structural diversity index $\Delta(G)$ by simulating $10^4$ realizations of $M_G$ and taking the average. Error bars are smaller than the sizes of symbols.

\bmhead{Figure \ref{fig:SFWS}} Simulations were performed on scale-free networks with $N = 10^3$ vertices and minimum degree $m=4$ and on Watts-Strogatz networks with $N= 10^3$ vertices and average degree $\langle k \rangle= 6$. 
We computed the structural diversity index $\Delta(G)$ by simulating $10^4$ realizations of $M_G$ and averaging. 
Error bars are smaller than the size of the symbols. 
The parameters $a$ and $b$ in (A) were obtained by an ordinary least squares fit of $\log(\Delta(G_{\gamma}))$ against $\log(\lvert V(G_{\gamma}) \rvert^{-(3-\gamma)/(\gamma-1)})$. 
The fit yields $a = 0.3$ and $b=6.2$ with $R^2=0.91$.

\bmhead{Figure \ref{fig:gprops}} For each real-world network $G$ in our network dataset (see Table B.1 and Table B.2 of the SI for details), we calculated (i) the structural diversity index, (ii) degree-heterogeneity, (iii) Wiener index, (iv) edge density, (v) clustering, and (vi) network size. 
We provide definitions of these quantities in the text and in Section B of the SI. 
We computed the structural diversity index and Wiener index in the same way as outlined for Figure \ref{fig:duality}; clustering using algorithms from graph-tool \cite{peixoto2014graphtool}; 
degree-heterogeneity, edge density, and size by evaluating simple mathematical expressions.

\bmhead{Figure \ref{fig:edge_removal}} For selected real-world networks $G$ in our network dataset (see Table B.1 and Table B.2 of the SI for details) and each $h=1,2,3,4$, we obtained the network $G_h$ through the edge removal procedure described in Figure \ref{fig:edge_removal}A. 
Specifically, $G_h$ is the largest connected component of the network obtained by the procedure in Figure \ref{fig:edge_removal}A. 
For each network $G_h$, we computed the structural diversity index by simulating $10^4$ realizations of $M_{G_h}$ and averaging them. 
Before creating the plot in Figure \ref{fig:edge_removal}B, we cleaned the data by discarding some ``pathological cases''. First, we discarded all networks with $\lvert V(G_h) \rvert < \lvert V(G)\rvert/4$. 
These networks were too affected by edge removal for comparisons to be meaningful. Second, we filtered out outliers, i.e., networks such that the relative variation of the structural diversity index $(\Delta(G_h)-\Delta(G))/\Delta(G)$ deviated more than $1$ standard deviation from the average relative variation of the sample.
This procedure discards just a few (about $4$) networks for each value of $h$. 
The discarded networks all follow a common pattern: the relative variations in their structural diversity indices are anomalously large because of specific structural features. 
Such anomalous cases are not interesting for our statistical study. 
This is why we discarded them. 
The red line is fitted using the ordinary least-squares method ($a = 0.105, R^2=0.223$).

\bmhead{Table \ref{table:regression}}
The coefficients, standard errors and $p$-values displayed in Table \ref{table:regression} are obtained by running ordinary least-square regressions on log-transformed standardized data. Due to space limitations, we excluded the regression analysis of structural diversity against size, as it was considered the least relevant. The network sample used in the regression is the same as that used in Figure \ref{fig:gprops}. 

\bmhead{Computing realizations of $M_G$}
A realization of $M_G$ is computed by simulating two random walks on the graph $G$. The simulation is run for $s_{max} = 100 \cdot \lvert V(G) \rvert$ steps. 
If the random walks do not meet within $s_{max}$ steps, we estimate the value of $M_G$. 
For this, we use the fact that $M_G$ is approximately geometrically distributed (see Section A of the SI). 
Specifically, we estimate the geometric distribution that best approximates $M_G$. 
Then we sample from this geometric distribution conditioned on the fact that the sampled value should exceed $s_{max}$.

\backmatter
\section*{Declarations}
\bmhead{Supplementary information}
This article is accompanied by a Supplementary Information (SI). 
\bmhead{Acknowledgments}
A.M. is grateful to Thomas Asikis for support in running simulations. We thank the two anonymous referees for their constructive feedback. This project received financial support from the European
Research Council (ERC) under the European Union’s Horizon 2020 research567
and innovation program (grant agreement No 833168).

\bmhead{Funding} This project received financial support from the European Research Council (ERC) under the European Union’s Horizon 2020 research and innovation program (grant agreement No 833168).
\bmhead{Competing interests} The authors declare no competing interests.
\bmhead{Ethics approval} Not applicable.
\bmhead{Consent to participate} Not applicable.
\bmhead{Consent to publication} Not applicable.
\bmhead{Availability of data and materials} The network data employed in this study are freely available at \url{https://networks.skewed.de}. 
\bmhead{Code availability} A Python package enabling the fast numerical computation of the structural diversity index developed by the first author is available through PyPi \url{https://pypi.org/project/structural-diversity-index/}.
A detailed tutorial on how to use the scripts is available on the \emph{eth-coss} public GitHub repository (see \url{https://github.com/ethz-coss/Structural-diversity-index}) and the full documentation of the scripts can be found on ReadTheDocs (see \url{https://rse-distance.readthedocs.io/en/latest/}).
\bmhead{Author contributions}
D.H. and A.M. designed research and wrote the paper; A.M. performed research, implemented simulation code, and contributed new analytic results.
\newpage
\nolinenumbers
\bibliography{sn-bibliography}


\begin{thebibliography}{65}
\ifx \bisbn   \undefined \def \bisbn  #1{ISBN #1}\fi
\ifx \binits  \undefined \def \binits#1{#1}\fi
\ifx \bauthor  \undefined \def \bauthor#1{#1}\fi
\ifx \batitle  \undefined \def \batitle#1{#1}\fi
\ifx \bjtitle  \undefined \def \bjtitle#1{#1}\fi
\ifx \bvolume  \undefined \def \bvolume#1{\textbf{#1}}\fi
\ifx \byear  \undefined \def \byear#1{#1}\fi
\ifx \bissue  \undefined \def \bissue#1{#1}\fi
\ifx \bfpage  \undefined \def \bfpage#1{#1}\fi
\ifx \blpage  \undefined \def \blpage #1{#1}\fi
\ifx \burl  \undefined \def \burl#1{\textsf{#1}}\fi
\ifx \doiurl  \undefined \def \doiurl#1{\url{https://doi.org/#1}}\fi
\ifx \betal  \undefined \def \betal{\textit{et al.}}\fi
\ifx \binstitute  \undefined \def \binstitute#1{#1}\fi
\ifx \binstitutionaled  \undefined \def \binstitutionaled#1{#1}\fi
\ifx \bctitle  \undefined \def \bctitle#1{#1}\fi
\ifx \beditor  \undefined \def \beditor#1{#1}\fi
\ifx \bpublisher  \undefined \def \bpublisher#1{#1}\fi
\ifx \bbtitle  \undefined \def \bbtitle#1{#1}\fi
\ifx \bedition  \undefined \def \bedition#1{#1}\fi
\ifx \bseriesno  \undefined \def \bseriesno#1{#1}\fi
\ifx \blocation  \undefined \def \blocation#1{#1}\fi
\ifx \bsertitle  \undefined \def \bsertitle#1{#1}\fi
\ifx \bsnm \undefined \def \bsnm#1{#1}\fi
\ifx \bsuffix \undefined \def \bsuffix#1{#1}\fi
\ifx \bparticle \undefined \def \bparticle#1{#1}\fi
\ifx \barticle \undefined \def \barticle#1{#1}\fi
\bibcommenthead
\ifx \bconfdate \undefined \def \bconfdate #1{#1}\fi
\ifx \botherref \undefined \def \botherref #1{#1}\fi
\ifx \url \undefined \def \url#1{\textsf{#1}}\fi
\ifx \bchapter \undefined \def \bchapter#1{#1}\fi
\ifx \bbook \undefined \def \bbook#1{#1}\fi
\ifx \bcomment \undefined \def \bcomment#1{#1}\fi
\ifx \oauthor \undefined \def \oauthor#1{#1}\fi
\ifx \citeauthoryear \undefined \def \citeauthoryear#1{#1}\fi
\ifx \endbibitem  \undefined \def \endbibitem {}\fi
\ifx \bconflocation  \undefined \def \bconflocation#1{#1}\fi
\ifx \arxivurl  \undefined \def \arxivurl#1{\textsf{#1}}\fi
\csname PreBibitemsHook\endcsname

\bibitem{dawkins1990selfish}
\begin{bbook}
\bauthor{\bsnm{Dawkins}, \binits{R.}}:
\bbtitle{The {{Selfish Gene}}},
\bedition{2nd edition} edn.
\bpublisher{{Oxford University Press}},
\blocation{{Oxford}}
(\byear{1990})
\end{bbook}
\endbibitem

\bibitem{nowak2006evolutionary}
\begin{bbook}
\bauthor{\bsnm{Nowak}, \binits{M.A.}}:
\bbtitle{Evolutionary {{Dynamics}}: {{Exploring}} the {{Equations}} of {{Life}}}.
\bpublisher{{Belknap Press}},
\blocation{{Cambridge}}
(\byear{2006})
\end{bbook}
\endbibitem

\bibitem{lieberman2005evolutionary}
\begin{barticle}
\bauthor{\bsnm{Lieberman}, \binits{E.}},
\bauthor{\bsnm{Hauert}, \binits{C.}},
\bauthor{\bsnm{Nowak}, \binits{M.A.}}:
\batitle{Evolutionary dynamics on graphs}.
\bjtitle{Nature}
\bvolume{433}(\bissue{7023}),
\bfpage{312}--\blpage{316}
(\byear{2005}).
\doiurl{10.1038/nature03204}
\end{barticle}
\endbibitem

\bibitem{bastolla2009architecture}
\begin{barticle}
\bauthor{\bsnm{Bastolla}, \binits{U.}},
\bauthor{\bsnm{Fortuna}, \binits{M.A.}},
\bauthor{\bsnm{{Pascual-Garc{\'i}a}}, \binits{A.}},
\bauthor{\bsnm{Ferrera}, \binits{A.}},
\bauthor{\bsnm{Luque}, \binits{B.}},
\bauthor{\bsnm{Bascompte}, \binits{J.}}:
\batitle{The architecture of mutualistic networks minimizes competition and increases biodiversity}.
\bjtitle{Nature}
\bvolume{458}(\bissue{7241}),
\bfpage{1018}--\blpage{1020}
(\byear{2009}).
\doiurl{10.1038/nature07950}
\end{barticle}
\endbibitem

\bibitem{boyd1988culture}
\begin{bbook}
\bauthor{\bsnm{Boyd}, \binits{R.}},
\bauthor{\bsnm{Richerson}, \binits{P.J.}}:
\bbtitle{Culture and the {{Evolutionary Process}}}.
\bpublisher{{University of Chicago Press}},
\blocation{{Chicago}}
(\byear{1988})
\end{bbook}
\endbibitem

\bibitem{simmel1957fashion}
\begin{barticle}
\bauthor{\bsnm{Simmel}, \binits{G.}}:
\batitle{Fashion}.
\bjtitle{American Journal of Sociology}
\bvolume{62}(\bissue{6}),
\bfpage{541}--\blpage{558}
(\byear{1957}).
\doiurl{10.1086/222102}
\end{barticle}
\endbibitem

\bibitem{granovetter1985economic}
\begin{barticle}
\bauthor{\bsnm{Granovetter}, \binits{M.}}:
\batitle{Economic {{Action}} and {{Social Structure}}: {{The Problem}} of {{Embeddedness}}}.
\bjtitle{American Journal of Sociology}
\bvolume{91}(\bissue{3}),
\bfpage{481}--\blpage{510}
(\byear{1985})
{\href{https://arxiv.org/abs/2780199}{{2780199}}}
\end{barticle}
\endbibitem

\bibitem{christakis2007spread}
\begin{barticle}
\bauthor{\bsnm{Christakis}, \binits{N.A.}},
\bauthor{\bsnm{Fowler}, \binits{J.H.}}:
\batitle{The {{Spread}} of {{Obesity}} in a {{Large Social Network}} over 32 {{Years}}}.
\bjtitle{New England Journal of Medicine}
\bvolume{357}(\bissue{4}),
\bfpage{370}--\blpage{379}
(\byear{2007}).
\doiurl{10.1056/NEJMsa066082}
\end{barticle}
\endbibitem

\bibitem{christakis2008collective}
\begin{barticle}
\bauthor{\bsnm{Christakis}, \binits{N.A.}},
\bauthor{\bsnm{Fowler}, \binits{J.H.}}:
\batitle{The {{Collective Dynamics}} of {{Smoking}} in a {{Large Social Network}}}.
\bjtitle{New England Journal of Medicine}
\bvolume{358}(\bissue{21}),
\bfpage{2249}--\blpage{2258}
(\byear{2008}).
\doiurl{10.1056/NEJMsa0706154}
\end{barticle}
\endbibitem

\bibitem{hauert2004spatial}
\begin{barticle}
\bauthor{\bsnm{Hauert}, \binits{C.}},
\bauthor{\bsnm{Doebeli}, \binits{M.}}:
\batitle{Spatial structure often inhibits the evolution of cooperation in the snowdrift game}.
\bjtitle{Nature}
\bvolume{428}(\bissue{6983}),
\bfpage{643}--\blpage{646}
(\byear{2004}).
\doiurl{10.1038/nature02360}
\end{barticle}
\endbibitem

\bibitem{nowak2006five}
\begin{barticle}
\bauthor{\bsnm{Nowak}, \binits{M.A.}}:
\batitle{Five {{Rules}} for the {{Evolution}} of {{Cooperation}}}.
\bjtitle{Science}
\bvolume{314}(\bissue{5805}),
\bfpage{1560}--\blpage{1563}
(\byear{2006}).
\doiurl{10.1126/science.1133755}
\end{barticle}
\endbibitem

\bibitem{granovetter1978threshold}
\begin{barticle}
\bauthor{\bsnm{Granovetter}, \binits{M.}}:
\batitle{Threshold {{Models}} of {{Collective Behavior}}}.
\bjtitle{American Journal of Sociology}
\bvolume{83}(\bissue{6}),
\bfpage{1420}--\blpage{1443}
(\byear{1978}).
\doiurl{10.1086/226707}
\end{barticle}
\endbibitem

\bibitem{centola2010spreada}
\begin{barticle}
\bauthor{\bsnm{Centola}, \binits{D.}}:
\batitle{The {{Spread}} of {{Behavior}} in an {{Online Social Network Experiment}}}.
\bjtitle{Science}
\bvolume{329}(\bissue{5996}),
\bfpage{1194}--\blpage{1197}
(\byear{2010}).
\doiurl{10.1126/science.1185231}
\end{barticle}
\endbibitem

\bibitem{ugander2012structurala}
\begin{barticle}
\bauthor{\bsnm{Ugander}, \binits{J.}},
\bauthor{\bsnm{Backstrom}, \binits{L.}},
\bauthor{\bsnm{Marlow}, \binits{C.}},
\bauthor{\bsnm{Kleinberg}, \binits{J.}}:
\batitle{Structural diversity in social contagion}.
\bjtitle{Proceedings of the National Academy of Sciences}
\bvolume{109}(\bissue{16}),
\bfpage{5962}--\blpage{5966}
(\byear{2012}).
\doiurl{10.1073/pnas.1116502109}
\end{barticle}
\endbibitem

\bibitem{onnela2007structure}
\begin{barticle}
\bauthor{\bsnm{Onnela}, \binits{J.-P.}},
\bauthor{\bsnm{Saram{\"a}ki}, \binits{J.}},
\bauthor{\bsnm{Hyv{\"o}nen}, \binits{J.}},
\bauthor{\bsnm{Szab{\'o}}, \binits{G.}},
\bauthor{\bsnm{Lazer}, \binits{D.}},
\bauthor{\bsnm{Kaski}, \binits{K.}},
\bauthor{\bsnm{Kert{\'e}sz}, \binits{J.}},
\bauthor{\bsnm{Barab{\'a}si}, \binits{A.-L.}}:
\batitle{Structure and tie strengths in mobile communication networks}.
\bjtitle{Proceedings of the National Academy of Sciences}
\bvolume{104}(\bissue{18}),
\bfpage{7332}--\blpage{7336}
(\byear{2007}).
\doiurl{10.1073/pnas.0610245104}
\end{barticle}
\endbibitem

\bibitem{granovetter1973strength}
\begin{barticle}
\bauthor{\bsnm{Granovetter}, \binits{M.S.}}:
\batitle{The {{Strength}} of {{Weak Ties}}}.
\bjtitle{American Journal of Sociology}
\bvolume{78}(\bissue{6}),
\bfpage{1360}--\blpage{1380}
(\byear{1973})
{\href{https://arxiv.org/abs/2776392}{{2776392}}}
\end{barticle}
\endbibitem

\bibitem{lazer2007network}
\begin{barticle}
\bauthor{\bsnm{Lazer}, \binits{D.}},
\bauthor{\bsnm{Friedman}, \binits{A.}}:
\batitle{The {{Network Structure}} of {{Exploration}} and {{Exploitation}}}.
\bjtitle{Administrative Science Quarterly}
\bvolume{52}(\bissue{4}),
\bfpage{667}--\blpage{694}
(\byear{2007}).
\doiurl{10.2189/asqu.52.4.667}
\end{barticle}
\endbibitem

\bibitem{mason2012collaborative}
\begin{barticle}
\bauthor{\bsnm{Mason}, \binits{W.}},
\bauthor{\bsnm{Watts}, \binits{D.J.}}:
\batitle{Collaborative learning in networks}.
\bjtitle{Proceedings of the National Academy of Sciences}
\bvolume{109}(\bissue{3}),
\bfpage{764}--\blpage{769}
(\byear{2012}).
\doiurl{10.1073/pnas.1110069108}
\end{barticle}
\endbibitem

\bibitem{rayleigh1905problem}
\begin{barticle}
\bauthor{\bsnm{Rayleigh}}:
\batitle{The {{Problem}} of the {{Random Walk}}}.
\bjtitle{Nature}
\bvolume{72}(\bissue{1866}),
\bfpage{318}--\blpage{318}
(\byear{1905}).
\doiurl{10.1038/072318a0}
\end{barticle}
\endbibitem

\bibitem{levin2017markov}
\begin{bbook}
\bauthor{\bsnm{Levin}, \binits{D.A.}},
\bauthor{\bsnm{Peres}, \binits{Y.}}:
\bbtitle{Markov {{Chains}} and {{Mixing Times}}}.
\bpublisher{{American Mathematical Society}},
\blocation{{Boston}}
(\byear{2017})
\end{bbook}
\endbibitem

\bibitem{appadurai1996modernity}
\begin{bbook}
\bauthor{\bsnm{Appadurai}, \binits{A.}}:
\bbtitle{Modernity {{At Large}}: {{Cultural Dimensions}} of {{Globalization}}}.
\bpublisher{{University of Minnesota Press}},
\blocation{{Minneapolis}}
(\byear{1996})
\end{bbook}
\endbibitem

\bibitem{barabasi1999emergence}
\begin{barticle}
\bauthor{\bsnm{Barab{\'a}si}, \binits{A.-L.}},
\bauthor{\bsnm{Albert}, \binits{R.}}:
\batitle{Emergence of {{Scaling}} in {{Random Networks}}}.
\bjtitle{Science}
\bvolume{286}(\bissue{5439}),
\bfpage{509}--\blpage{512}
(\byear{1999}).
\doiurl{10.1126/science.286.5439.509}
\end{barticle}
\endbibitem

\bibitem{watts1998collective}
\begin{barticle}
\bauthor{\bsnm{Watts}, \binits{D.J.}},
\bauthor{\bsnm{Strogatz}, \binits{S.H.}}:
\batitle{Collective dynamics of `small-world' networks}.
\bjtitle{Nature}
\bvolume{393}(\bissue{6684}),
\bfpage{440}--\blpage{442}
(\byear{1998}).
\doiurl{10.1038/30918}
\end{barticle}
\endbibitem

\bibitem{peixoto2014graphtool}
\begin{botherref}
\oauthor{\bsnm{Peixoto}, \binits{T.}}:
The Graph-Tool Python Library.
{figshare}
(2014).
\doiurl{10.6084/m9.figshare.1164194.v14}
\end{botherref}
\endbibitem

\bibitem{axelrod1997disseminationa}
\begin{barticle}
\bauthor{\bsnm{Axelrod}, \binits{R.}}:
\batitle{The {{Dissemination}} of {{Culture}}: {{A Model}} with {{Local Convergence}} and {{Global Polarization}}}.
\bjtitle{Journal of Conflict Resolution}
\bvolume{41}(\bissue{2}),
\bfpage{203}--\blpage{226}
(\byear{1997}).
\doiurl{10.1177/0022002797041002001}
\end{barticle}
\endbibitem

\bibitem{weitzman1992diversity}
\begin{barticle}
\bauthor{\bsnm{Weitzman}, \binits{M.L.}}:
\batitle{On {{Diversity}}*}.
\bjtitle{The Quarterly Journal of Economics}
\bvolume{107}(\bissue{2}),
\bfpage{363}--\blpage{405}
(\byear{1992}).
\doiurl{10.2307/2118476}
\end{barticle}
\endbibitem

\bibitem{huckfeldt2004political}
\begin{bbook}
\bauthor{\bsnm{Huckfeldt}, \binits{R.R.}},
\bauthor{\bsnm{Johnson}, \binits{P.E.}},
\bauthor{\bsnm{Sprague}, \binits{J.D.}}:
\bbtitle{Political {{Disagreement}}: {{The Survival}} of {{Diverse Opinions Within Communication Networks}}}.
\bpublisher{{Cambridge University Press}},
\blocation{{Cambridge}}
(\byear{2004})
\end{bbook}
\endbibitem

\bibitem{stirling2007general}
\begin{barticle}
\bauthor{\bsnm{Stirling}, \binits{A.}}:
\batitle{A general framework for analysing diversity in science, technology and society}.
\bjtitle{Journal of The Royal Society Interface}
\bvolume{4}(\bissue{15}),
\bfpage{707}--\blpage{719}
(\byear{2007}).
\doiurl{10.1098/rsif.2007.0213}
\end{barticle}
\endbibitem

\bibitem{klemm2003global}
\begin{barticle}
\bauthor{\bsnm{Klemm}, \binits{K.}},
\bauthor{\bsnm{Egu{\'i}luz}, \binits{V.M.}},
\bauthor{\bsnm{Toral}, \binits{R.}},
\bauthor{\bsnm{Miguel}, \binits{M.S.}}:
\batitle{Global culture: {{A}} noise-induced transition in finite systems}.
\bjtitle{Physical Review E}
\bvolume{67}(\bissue{4}),
\bfpage{045101}
(\byear{2003}).
\doiurl{10.1103/PhysRevE.67.045101}
\end{barticle}
\endbibitem

\bibitem{page2008difference}
\begin{bbook}
\bauthor{\bsnm{Page}, \binits{S.E.}}:
\bbtitle{The {{Difference}}}.
\bpublisher{{Princeton University Press}},
\blocation{{Princeton}}
(\byear{Sun, 08/31/2008 - 12:00})
\end{bbook}
\endbibitem

\bibitem{feldman1999innovation}
\begin{barticle}
\bauthor{\bsnm{Feldman}, \binits{M.P.}},
\bauthor{\bsnm{Audretsch}, \binits{D.B.}}:
\batitle{Innovation in cities:: {{Science-based}} diversity, specialization and localized competition}.
\bjtitle{European Economic Review}
\bvolume{43}(\bissue{2}),
\bfpage{409}--\blpage{429}
(\byear{1999}).
\doiurl{10.1016/S0014-2921(98)00047-6}
\end{barticle}
\endbibitem

\bibitem{santos2008social}
\begin{barticle}
\bauthor{\bsnm{Santos}, \binits{F.C.}},
\bauthor{\bsnm{Santos}, \binits{M.D.}},
\bauthor{\bsnm{Pacheco}, \binits{J.M.}}:
\batitle{Social diversity promotes the emergence of cooperation in public goods games}.
\bjtitle{Nature}
\bvolume{454}(\bissue{7201}),
\bfpage{213}--\blpage{216}
(\byear{2008}).
\doiurl{10.1038/nature06940}
\end{barticle}
\endbibitem

\bibitem{vasconcelos2021segregationa}
\begin{barticle}
\bauthor{\bsnm{Vasconcelos}, \binits{V.V.}},
\bauthor{\bsnm{Constantino}, \binits{S.M.}},
\bauthor{\bsnm{Dannenberg}, \binits{A.}},
\bauthor{\bsnm{Lumkowsky}, \binits{M.}},
\bauthor{\bsnm{Weber}, \binits{E.}},
\bauthor{\bsnm{Levin}, \binits{S.}}:
\batitle{Segregation and clustering of preferences erode socially beneficial coordination}.
\bjtitle{Proceedings of the National Academy of Sciences}
\bvolume{118}(\bissue{50}),
\bfpage{2102153118}
(\byear{2021}).
\doiurl{10.1073/pnas.2102153118}
\end{barticle}
\endbibitem

\bibitem{bettencourt2014professional}
\begin{barticle}
\bauthor{\bsnm{Bettencourt}, \binits{L.M.A.}},
\bauthor{\bsnm{Samaniego}, \binits{H.}},
\bauthor{\bsnm{Youn}, \binits{H.}}:
\batitle{Professional diversity and the productivity of cities}.
\bjtitle{Scientific Reports}
\bvolume{4}(\bissue{1}),
\bfpage{5393}
(\byear{2014}).
\doiurl{10.1038/srep05393}
\end{barticle}
\endbibitem

\bibitem{gomez-lievano2016explaininga}
\begin{barticle}
\bauthor{\bsnm{{Gomez-Lievano}}, \binits{A.}},
\bauthor{\bsnm{{Patterson-Lomba}}, \binits{O.}},
\bauthor{\bsnm{Hausmann}, \binits{R.}}:
\batitle{Explaining the prevalence, scaling and variance of urban phenomena}.
\bjtitle{Nature Human Behaviour}
\bvolume{1}(\bissue{1}),
\bfpage{1}--\blpage{6}
(\byear{2016}).
\doiurl{10.1038/s41562-016-0012}
\end{barticle}
\endbibitem

\bibitem{centola2022network}
\begin{barticle}
\bauthor{\bsnm{Centola}, \binits{D.}}:
\batitle{The network science of collective intelligence}.
\bjtitle{Trends in Cognitive Sciences}
\bvolume{26}(\bissue{11}),
\bfpage{923}--\blpage{941}
(\byear{2022}).
\doiurl{10.1016/j.tics.2022.08.009}
\end{barticle}
\endbibitem

\bibitem{hong2004groups}
\begin{barticle}
\bauthor{\bsnm{Hong}, \binits{L.}},
\bauthor{\bsnm{Page}, \binits{S.E.}}:
\batitle{Groups of diverse problem solvers can outperform groups of high-ability problem solvers}.
\bjtitle{Proceedings of the National Academy of Sciences}
\bvolume{101}(\bissue{46}),
\bfpage{16385}--\blpage{16389}
(\byear{2004}).
\doiurl{10.1073/pnas.0403723101}
\end{barticle}
\endbibitem

\bibitem{lorenz2011howa}
\begin{barticle}
\bauthor{\bsnm{Lorenz}, \binits{J.}},
\bauthor{\bsnm{Rauhut}, \binits{H.}},
\bauthor{\bsnm{Schweitzer}, \binits{F.}},
\bauthor{\bsnm{Helbing}, \binits{D.}}:
\batitle{How social influence can undermine the wisdom of crowd effect}.
\bjtitle{Proceedings of the National Academy of Sciences}
\bvolume{108}(\bissue{22}),
\bfpage{9020}--\blpage{9025}
(\byear{2011}).
\doiurl{10.1073/pnas.1008636108}
\end{barticle}
\endbibitem

\bibitem{bernstein2018how}
\begin{barticle}
\bauthor{\bsnm{Bernstein}, \binits{E.}},
\bauthor{\bsnm{Shore}, \binits{J.}},
\bauthor{\bsnm{Lazer}, \binits{D.}}:
\batitle{How intermittent breaks in interaction improve collective intelligence}.
\bjtitle{Proceedings of the National Academy of Sciences}
\bvolume{115}(\bissue{35}),
\bfpage{8734}--\blpage{8739}
(\byear{2018}).
\doiurl{10.1073/pnas.1802407115}
\end{barticle}
\endbibitem

\bibitem{woolley2010evidence}
\begin{barticle}
\bauthor{\bsnm{Woolley}, \binits{A.W.}},
\bauthor{\bsnm{Chabris}, \binits{C.F.}},
\bauthor{\bsnm{Pentland}, \binits{A.}},
\bauthor{\bsnm{Hashmi}, \binits{N.}},
\bauthor{\bsnm{Malone}, \binits{T.W.}}:
\batitle{Evidence for a {{Collective Intelligence Factor}} in the {{Performance}} of {{Human Groups}}}.
\bjtitle{Science}
\bvolume{330}(\bissue{6004}),
\bfpage{686}--\blpage{688}
(\byear{2010}).
\doiurl{10.1126/science.1193147}
\end{barticle}
\endbibitem

\bibitem{guzzo1996teams}
\begin{barticle}
\bauthor{\bsnm{Guzzo}, \binits{R.A.}},
\bauthor{\bsnm{Dickson}, \binits{M.W.}}:
\batitle{Teams in organizations: {{Recent}} research on performance and effectiveness}.
\bjtitle{Annual Review of Psychology}
\bvolume{47},
\bfpage{307}--\blpage{338}
(\byear{1996}).
\doiurl{10.1146/annurev.psych.47.1.307}
\end{barticle}
\endbibitem

\bibitem{webber2001impact}
\begin{barticle}
\bauthor{\bsnm{Webber}, \binits{S.S.}},
\bauthor{\bsnm{Donahue}, \binits{L.M.}}:
\batitle{Impact of highly and less job-related diversity on work group cohesion and performance: A meta-analysis}.
\bjtitle{Journal of Management}
\bvolume{27}(\bissue{2}),
\bfpage{141}--\blpage{162}
(\byear{2001}).
\doiurl{10.1177/014920630102700202}
\end{barticle}
\endbibitem

\bibitem{zenger1989organizational}
\begin{barticle}
\bauthor{\bsnm{Zenger}, \binits{T.R.}},
\bauthor{\bsnm{Lawrence}, \binits{B.S.}}:
\batitle{Organizational demography: {{The}} differential effects of age and tenure distributions on technical communication}.
\bjtitle{Academy of Management Journal}
\bvolume{32}(\bissue{2}),
\bfpage{353}--\blpage{376}
(\byear{1989}).
\doiurl{10.2307/256366}
\end{barticle}
\endbibitem

\bibitem{glaeser2000measuring}
\begin{barticle}
\bauthor{\bsnm{Glaeser}, \binits{E.L.}},
\bauthor{\bsnm{Laibson}, \binits{D.I.}},
\bauthor{\bsnm{Scheinkman}, \binits{J.A.}},
\bauthor{\bsnm{Soutter}, \binits{C.L.}}:
\batitle{Measuring {{Trust}}*}.
\bjtitle{The Quarterly Journal of Economics}
\bvolume{115}(\bissue{3}),
\bfpage{811}--\blpage{846}
(\byear{2000}).
\doiurl{10.1162/003355300554926}
\end{barticle}
\endbibitem

\bibitem{putnam2007pluribus}
\begin{barticle}
\bauthor{\bsnm{Putnam}, \binits{R.D.}}:
\batitle{E {{Pluribus Unum}}: {{Diversity}} and {{Community}} in the {{Twenty-first Century The}} 2006 {{Johan Skytte Prize Lecture}}}.
\bjtitle{Scandinavian Political Studies}
\bvolume{30}(\bissue{2}),
\bfpage{137}--\blpage{174}
(\byear{2007}).
\doiurl{10.1111/j.1467-9477.2007.00176.x}
\end{barticle}
\endbibitem

\bibitem{milgram1967small}
\begin{barticle}
\bauthor{\bsnm{Milgram}, \binits{S.}}:
\batitle{The small world problem}.
\bjtitle{Psychology Today}
\bvolume{2},
\bfpage{60}--\blpage{67}
(\byear{1967})
\end{barticle}
\endbibitem

\bibitem{aldous1991meeting}
\begin{barticle}
\bauthor{\bsnm{Aldous}, \binits{D.J.}}:
\batitle{Meeting times for independent {{Markov}} chains}.
\bjtitle{Stochastic Processes and their Applications}
\bvolume{38}(\bissue{2}),
\bfpage{185}--\blpage{193}
(\byear{1991}).
\doiurl{10.1016/0304-4149(91)90090-Y}
\end{barticle}
\endbibitem

\bibitem{kanade2023coalescence}
\begin{barticle}
\bauthor{\bsnm{Kanade}, \binits{V.}},
\bauthor{\bsnm{{Mallmann-Trenn}}, \binits{F.}},
\bauthor{\bsnm{Sauerwald}, \binits{T.}}:
\batitle{On {{Coalescence Time}} in {{Graphs}}: {{When Is Coalescing}} as {{Fast}} as {{Meeting}}?}
\bjtitle{ACM Transactions on Algorithms}
\bvolume{19}(\bissue{2}),
\bfpage{18}--\blpage{11846}
(\byear{2023}).
\doiurl{10.1145/3576900}
\end{barticle}
\endbibitem

\bibitem{liggett1985voter}
\begin{bchapter}
\bauthor{\bsnm{Liggett}, \binits{T.M.}}:
\bctitle{The {{Voter Model}}}.
In: \beditor{\bsnm{Liggett}, \binits{T.M.}} (ed.)
\bbtitle{Interacting {{Particle Systems}}}.
\bsertitle{Grundlehren Der Mathematischen {{Wissenschaften}}},
pp. \bfpage{226}--\blpage{263}.
\bpublisher{{Springer}},
\blocation{{New York, NY}}
(\byear{1985}).
\doiurl{10.1007/978-1-4613-8542-4_6}
\end{bchapter}
\endbibitem

\bibitem{simpson1949measurementa}
\begin{barticle}
\bauthor{\bsnm{Simpson}, \binits{E.H.}}:
\batitle{Measurement of {{Diversity}}}.
\bjtitle{Nature}
\bvolume{163}(\bissue{4148}),
\bfpage{688}--\blpage{688}
(\byear{1949}).
\doiurl{10.1038/163688a0}
\end{barticle}
\endbibitem

\bibitem{aldous2013probability}
\begin{bbook}
\bauthor{\bsnm{Aldous}, \binits{D.}}:
\bbtitle{Probability {{Approximations}} Via the {{Poisson Clumping Heuristic}}}.
\bpublisher{{Springer Science \& Business Media}},
\blocation{{Berlin}}
(\byear{2013})
\end{bbook}
\endbibitem

\bibitem{barrat2008dynamical}
\begin{bbook}
\bauthor{\bsnm{Barrat}, \binits{A.}},
\bauthor{\bsnm{Barth{\'e}lemy}, \binits{M.}},
\bauthor{\bsnm{Vespignani}, \binits{A.}}:
\bbtitle{Dynamical {{Processes}} on {{Complex Networks}}}.
\bpublisher{{Cambridge University Press}},
\blocation{{Cambridge}}
(\byear{2008}).
\doiurl{10.1017/CBO9780511791383}
\end{bbook}
\endbibitem

\bibitem{watts2007influentials}
\begin{barticle}
\bauthor{\bsnm{Watts}, \binits{D.J.}},
\bauthor{\bsnm{Dodds}, \binits{P.S.}}:
\batitle{Influentials, {{Networks}}, and {{Public Opinion Formation}}}.
\bjtitle{Journal of Consumer Research}
\bvolume{34}(\bissue{4}),
\bfpage{441}--\blpage{458}
(\byear{2007}).
\doiurl{10.1086/518527}
\end{barticle}
\endbibitem

\bibitem{borondo2014each}
\begin{barticle}
\bauthor{\bsnm{Borondo}, \binits{J.}},
\bauthor{\bsnm{Borondo}, \binits{F.}},
\bauthor{\bsnm{{Rodriguez-Sickert}}, \binits{C.}},
\bauthor{\bsnm{Hidalgo}, \binits{C.A.}}:
\batitle{To {{Each According}} to its {{Degree}}: {{The Meritocracy}} and {{Topocracy}} of {{Embedded Markets}}}.
\bjtitle{Scientific Reports}
\bvolume{4}(\bissue{1}),
\bfpage{3784}
(\byear{2014}).
\doiurl{10.1038/srep03784}
\end{barticle}
\endbibitem

\bibitem{sznajd-weron2000opinion}
\begin{barticle}
\bauthor{\bsnm{{Sznajd-Weron}}, \binits{K.}},
\bauthor{\bsnm{Sznajd}, \binits{J.}}:
\batitle{Opinion evolution in closed community}.
\bjtitle{International Journal of Modern Physics C}
\bvolume{11}(\bissue{06}),
\bfpage{1157}--\blpage{1165}
(\byear{2000}).
\doiurl{10.1142/S0129183100000936}
\end{barticle}
\endbibitem

\bibitem{hegselmann2005opinion}
\begin{barticle}
\bauthor{\bsnm{Hegselmann}, \binits{R.}},
\bauthor{\bsnm{Krause}, \binits{U.}}:
\batitle{Opinion {{Dynamics Driven}} by {{Various Ways}} of {{Averaging}}}.
\bjtitle{Computational Economics}
\bvolume{25}(\bissue{4}),
\bfpage{381}--\blpage{405}
(\byear{2005}).
\doiurl{10.1007/s10614-005-6296-3}
\end{barticle}
\endbibitem

\bibitem{lorenz2007continuous}
\begin{barticle}
\bauthor{\bsnm{Lorenz}, \binits{J.}}:
\batitle{Continuous opinion dynamics under bounded confidence: A survey}.
\bjtitle{International Journal of Modern Physics C}
\bvolume{18}(\bissue{12}),
\bfpage{1819}--\blpage{1838}
(\byear{2007}).
\doiurl{10.1142/S0129183107011789}
\end{barticle}
\endbibitem

\bibitem{castellano2009statisticala}
\begin{barticle}
\bauthor{\bsnm{Castellano}, \binits{C.}},
\bauthor{\bsnm{Fortunato}, \binits{S.}},
\bauthor{\bsnm{Loreto}, \binits{V.}}:
\batitle{Statistical physics of social dynamics}.
\bjtitle{Reviews of Modern Physics}
\bvolume{81}(\bissue{2}),
\bfpage{591}--\blpage{646}
(\byear{2009}).
\doiurl{10.1103/RevModPhys.81.591}
\end{barticle}
\endbibitem

\bibitem{livan2019don}
\begin{barticle}
\bauthor{\bsnm{Livan}, \binits{G.}}:
\batitle{Don't follow the leader: How ranking performance reduces meritocracy}.
\bjtitle{Royal Society Open Science}
\bvolume{6}(\bissue{11}),
\bfpage{191255}
(\byear{2019}).
\doiurl{10.1098/rsos.191255}
\end{barticle}
\endbibitem

\bibitem{pearson1905problem}
\begin{barticle}
\bauthor{\bsnm{Pearson}, \binits{K.}}:
\batitle{The {{Problem}} of the {{Random Walk}}}.
\bjtitle{Nature}
\bvolume{72}(\bissue{1865}),
\bfpage{294}--\blpage{294}
(\byear{1905}).
\doiurl{10.1038/072294b0}
\end{barticle}
\endbibitem

\bibitem{aldous2013interacting}
\begin{barticle}
\bauthor{\bsnm{Aldous}, \binits{D.}}:
\batitle{Interacting particle systems as stochastic social dynamics}.
\bjtitle{Bernoulli}
\bvolume{19}(\bissue{4}),
\bfpage{1122}--\blpage{1149}
(\byear{2013}).
\doiurl{10.3150/12-BEJSP04}
\end{barticle}
\endbibitem

\bibitem{cooper2013coalescing}
\begin{barticle}
\bauthor{\bsnm{Cooper}, \binits{C.}},
\bauthor{\bsnm{Els{\"a}sser}, \binits{R.}},
\bauthor{\bsnm{Ono}, \binits{H.}},
\bauthor{\bsnm{Radzik}, \binits{T.}}:
\batitle{Coalescing {{Random Walks}} and {{Voting}} on {{Connected Graphs}}}.
\bjtitle{SIAM Journal on Discrete Mathematics}
\bvolume{27}(\bissue{4}),
\bfpage{1748}--\blpage{1758}
(\byear{2013}).
\doiurl{10.1137/120900368}
\end{barticle}
\endbibitem

\bibitem{gkantsidis2003conductance}
\begin{bchapter}
\bauthor{\bsnm{Gkantsidis}, \binits{C.}},
\bauthor{\bsnm{Mihail}, \binits{M.}},
\bauthor{\bsnm{Saberi}, \binits{A.}}:
\bctitle{Conductance and congestion in power law graphs}.
In: \bbtitle{Proceedings of the 2003 {{ACM SIGMETRICS}} International Conference on {{Measurement}} and Modeling of Computer Systems}.
\bsertitle{{{SIGMETRICS}} '03},
pp. \bfpage{148}--\blpage{159}.
\bpublisher{{Association for Computing Machinery}},
\blocation{{New York, NY, USA}}
(\byear{2003}).
\doiurl{10.1145/781027.781046}
\end{bchapter}
\endbibitem

\bibitem{mihail2003certain}
\begin{bchapter}
\bauthor{\bsnm{Mihail}, \binits{M.}},
\bauthor{\bsnm{Papadimitriou}, \binits{C.}},
\bauthor{\bsnm{Saberi}, \binits{A.}}:
\bctitle{On certain connectivity properties of the {{Internet}} topology}.
In: \bbtitle{44th {{Annual IEEE Symposium}} on {{Foundations}} of {{Computer Science}}, 2003. {{Proceedings}}.},
pp. \bfpage{28}--\blpage{35}
(\byear{2003}).
\doiurl{10.1109/SFCS.2003.1238178}
\end{bchapter}
\endbibitem

\bibitem{hagberg2008exploring}
\begin{botherref}
\oauthor{\bsnm{Hagberg}, \binits{A.}},
\oauthor{\bsnm{Swart}, \binits{P.J.}},
\oauthor{\bsnm{Schult}, \binits{D.A.}}:
Exploring network structure, dynamics, and function using {{NetworkX}}.
Technical Report LA-UR-08-05495; LA-UR-08-5495,
{Los Alamos National Laboratory (LANL), Los Alamos, NM (United States)}
(2008)
\end{botherref}
\endbibitem

\end{thebibliography}


\end{document}